%% file: main.tex
\documentclass[sigconf]{acmart}

\usepackage{graphicx} % Required for inserting images
\usepackage{color}
\usepackage{xspace}
\usepackage{amsthm}
\usepackage[inline]{enumitem}
\usepackage{url}
\usepackage{footmisc}
\usepackage{tikz}
\usetikzlibrary{arrows.meta, positioning}
\usepackage[none]{hyphenat} % avoid word hyphenation

\input{macros}

\title{Domain Extension of Lock-Freedom and Wait-Freedom for Group Computations}
\author{Raaghav Ravishankar}\orcid{0009-0000-9918-4433}
\affiliation{
\institution{Michigan State University}
\city{East Lansing}
\country{USA}
}
\email{ravisha7@msu.edu}

\author{Sandeep Kulkarni}\orcid{0000-0002-7608-2116}
\affiliation{
\institution{Michigan State University}
\city{East Lansing}
\country{USA}
}
\email{sandeep@cse.msu.edu}

\author{Sathya Peri}\orcid{0000-0002-3471-7929}
\affiliation{
\institution{Indian Institute of Technology}
\city{Hyderabad}
\country{India}
}
\email{sathya_p@cse.iith.ac.in}

\author{Gokarna Sharma}\orcid{0000-0002-4930-4609}
\affiliation{
\institution{Kent State University}
\city{Kent}
\country{USA}
}
\email{gsharma2@kent.edu}

\author{Manaswini Piduguralla}\orcid{0000-0002-6295-0445}
\affiliation{
\institution{Indian Institute of Technology}
\city{Hyderabad}
\country{India}
}
\email{cs20resch11007@iith.ac.in}

\author{Yashi Rastogi}\orcid{0009-0000-3707-5004}
\affiliation{
\institution{Indian Institute of Technology}
\city{Hyderabad}
\country{India}
}
\email{cs23resch11007@iith.ac.in}

\begin{document}

\begin{abstract}

A domain extension\footnotemark of a definition refers to broadening the scope of a definition so that it applies to a larger set of cases than originally specified. 
The notion of lock-free and wait-free computation is designed for the domain of tasks that are completed by a single thread (in competition with other threads). The goal of this paper is to extend the definition of lock-freedom and wait-freedom to group-computations (denoted by \gl-freedom and \gw-freedom) that require that the task at hand \textit{must} be completed by a collaboration between multiple threads. 
When extending a definition, certain constraints must be respected: the new domain must remain logically consistent with the original meaning, the extension should not introduce contradictions or ambiguities, and it must preserve the essential properties that make the definition valid and useful. We demonstrate this by showing that our extended definition is consistent with the original definition when the group consists of a single thread. 
We note that extension allows us to characterize programs in a new domain (distributed computing, NUMA computation systems, systems with private data for different threads, etc.) instead of relegating them to be in the same category (deadlock/livelock-free) without regard to the actual properties of that program. We also illustrate this definition with various examples.

   % Progress guarantees such as lock freedom, and wait freedom have been foundational in concurrent computation, particularly within single-machine, multi-core environments. However, these definitions cannot be applied to multi-machine systems, where, out of necessity, operations span multiple collaborating threads and rely on system-level services. This paper addresses the limitations of classical progress models by introducing generalized definitions suitable for various distributed architectures. Central to our approach is the concept of \threadgroups (collections of cooperating threads) and \systemthreads, which assist multiple groups without belonging to any. We redefine wait-freedom and lock-freedom in terms of these groups.  We show that these new definitions preserve the semantics of lock-free and wait-free computation and can be realized in realistic system constraints. 
\end{abstract}

\maketitle

\section{Introduction}
\label{sec:intro}

% SPACE 
% A range of progress guarantees has been extensively studied within the domain of concurrent computation \cite{on-nature-of-progress,herlihy1991wait}. These guarantees, ordered by increasing strength, include deadlock freedom, starvation freedom, lock freedom, and wait freedom.
% Deadlock freedom ensures that at least one concurrent entity (typically a thread) is always able to complete its task. Starvation freedom strengthens this by requiring that every thread eventually completes its operation. Both of these guarantees generally presume the absence of thread failures, such as the failure of a thread holding a lock.
% To accommodate scenarios involving thread failure, lock freedom mandates that some thread can complete its operation even if others fail. Wait freedom further strengthens this guarantee by requiring that every thread can complete its operation regardless of delays, failures, or indefinite suspension of other threads.

%\spnote{The paper is not in the right format.}

Wait-freedom and lock-freedom are highly desirable properties \cite{herlihy1991wait,on-nature-of-progress} of a concurrent (multi-core) system. They offer strong guarantees of progress beyond deadlock-freedom and starvation-freedom. For example, wait-freedom guarantees that there is a worst-case response time (in terms of number of steps) for the operation.  
%This is critical in real-time systems \cite{Ermedahl98}. 
% \sandeepnote{We must find a citation for this.}

A key characteristic of these definitions is that they are thread-based, i.e., they assume that there is a (single) thread that invokes the operation and the same thread returns the result of that operation. 
%\spnote{With \wfdom it is not true that the same thread completes the \op} 
It may compete with other threads for resources or access of shared memory. However, inherently, the thread has the capability to complete the task on its own. At the bare minimum, it is expected that if the thread runs in isolation
%\raaghavnote{[Should we say "At least one thread will be able to complete the task" for a stronger constraint? Being able to run in isolation is only obstruction freedom.]}, 
it will complete the task even if the system contains no other threads.
%\sandeepnote{The point was to contradict with the next paragraph. I have made that more precise}

In this paper, we focus on the following question: \textit{What if the thread cannot complete the task on its own even if it runs in isolation, but rather, the task is collaborative and must be completed with a set of threads in a collaborative manner? }
%In this paper, we focus on the question: What if the task being conducted is a collaborative task, i.e., it must be completed by a set of threads in a collaborative manner. 
% \newtext{
One such example is a partitioned data structure with operations that span across two machines, $M1$ and $M2$. To complete an operation successfully, it would be necessary that at least one thread on $M1$ and at least one thread on $M2$ performs their task. A thread on $M1$ alone cannot complete the operation if it needs data from $M2$.
% } 

Another example is committee coordination with a set of principals with their own private calendar. The goal is to schedule different meetings (e.g., one meeting between $A$, $B$ and $C$ and another meeting between $A$, $C$ and $D$).
%between different set of principals. For example, one task is to schedule a meeting between A, B and C while another task is to schedule a meeting between A, C and D. If the calendar of each principal is private then this is a distributed collaboration problem where threads/processes of each principal must work together to schedule a meeting. 
One way to think about this is that there are several threads associated with each principal. The thread group A1, B1 and C1 is tasked with scheduling the meeting among the principals A, B and C while thread group A2, C2 and D2 is tasked with scheduling the meeting between A, C and D. Clearly, within a group, threads are \textit{collaborating}, and between groups, threads are \textit{competing}. 
%these groups are collaborating and completing \spnote{[should this be competing?]}. 
Specifically, A1 and A2 are competing to get a calendar slot for A, while A1, B1 and C1 are collaborating to find a common time slot. 

Here, we can observe that there is an inherent necessity for A1, B1 and C1 to collaborate and possibly wait for each other. In other words, if A1 fails or is slow, other threads cannot complete the task because they cannot perform the task associated with A1. On the other hand, if A1 fails or is slow, A2, C2 and D2 should be able to complete their task. %\spnote{Is the example of principals (with so much text) still relevant when the other example of lists seems more relevant. It is probably better to remove or shorten this example of principals}

Since the definition of lock-freedom and wait-freedom assumes that a thread is able to complete its task if it were not for the competition with other tasks, it does not distinguish between the algorithm  where A1 (respectively, A2) locks the entire calendar of A to perform the task or a more complicated algorithm where A2 (in conjunction with C2 and D2) can complete the task even if A1 fails or is slow. The goal of this paper is to extend the  spirit of lock-freedom (called \emph{\gl-freedom})  and wait-freedom (called \emph{\gw-freedom})  for these collaborative tasks. (A \gl-free solution to this problem is presented in 
the Appendix).
% \cite{ravishankar2026domainextensionlockfreedomwaitfreedomextended}).

\input{committeexample}

% Here is another simple instance of a problem that necessitates the extension of lock-freedom and wait-freedom. A server maintains a list, and a client accesses it via RPC for operations such as insert/delete/loookup. 
% This list will be classified only as deadlock-free without regard to whether the implementation locks the entire list or uses lock-free lists such as Harris \cite{Harris2001LinkedList}. {\bf GS:  confused. some elaboration may help}
% We believe that this lack of nuance does a disservice to understanding the characteristic of the algorithm. 
% %thereby losing a key distinction.  

%In this case, a client thread and a server thread must work together to complete this task. A client thread or a server thread alone cannot complete the task. Furthermore, this extension will allow us to distinguish between an implementation that uses Harris linked list \cite{?} and another lock based implementation

\textbf{Why Extend the Definition. }
Knowing that an algorithm is wait-free gives us a much stronger understanding of its behavior than simply knowing that it is deadlock free. Specifically, wait freedom guarantees that the system continues to make progress even if some \textit{threads} halt or fail.
However, for tasks designed to be carried out by groups of threads, current progress definitions allow us to classify them only as deadlock-free but not wait-free as threads in the same group can wait for each other. 
Identifying them instead as \gw-free (extension of wait-freedom for a group based computation) would tell us something more powerful: even if some \textit{groups} stop functioning, the remaining groups can still complete their \textit{task}. 
In this way, \gw-free is a stronger guarantee than deadlock-freedom for a group task, much like wait-freedom in the single-threaded task.
% In other words, identifying an algorithm as \gw-free provides the same kind of stronger guarantee that wait-freedom offers over deadlock-freedom.

% Extending the definition in this manner proves useful 
% In domains where threads participating in one task remain unaffected by threads engaged in another task, group guarantees provide substantive insight.
Extending this definition in this manner introduces a guarantee that threads of one task cannot interfere with threads of another task. 
%Extending the definition in this manner will be valuable in various domains where the knowledge that threads involved in one task cannot be affected by threads involved in another task will provide a valuable insight. 
As an example, 
%This issue also arises in several other domains.
consider a distributed data structure partitioned between two machines. To perform even a read-only operation that can span both machines, the initiator of the operation must inherently wait for threads from both machines to complete their portion of the task.
% Here, knowing that a thread
Likewise, matrix multiplication, numerical simulations or machine learning training on large datasets are split across machines due to the inherent limitation of putting all the data in one machine. For example, one thread decodes video, another handles audio, while another manages buffering and synchronization. The thread that is decoding the video may need to compete with other threads that are doing the same. But a thread that decodes the video is collaborating with another thread that handles the corresponding audio.
As we can see, here, there are some threads that are competing with each other and, hence, waiting between them is undesirable, whereas there are some threads that are collaborating with each other, and hence waiting between them is unavoidable.

\paragraph{Contributions of the paper}
% \raaghavnote{Revisit}
We introduce the notion of \threadgroups and present an extension of the lock-free and wait-free definitions (called \emph{\gl-free} and \emph{\gw-free}). We consider two variations of these definitions \strong and \weak.  We show that for single-threaded \threadgroups, both definitions coincide with the original formulation in \cite{herlihy1991wait,on-nature-of-progress}. We observe that while \strong \gl-freedom/\gw-freedom may be desirable, it is impossible for non-trivial group-based problems. Hence, we present \weak \gl-free solutions to three data structures: (1) protected list, (2) multi-machine list, and (3) distributed and partitioned list. We also investigate the space between \strong \gl/\gw-freedom and \weak \gl/\gw-freedom. We demonstrate that existing user-level software (ZeroMQ) and existing hardware support can be utilized to design \weak \gl/\gw-free algorithms without requiring kernel support. Based on this discussion, we argue that the \APT primitive presented in this paper is the foundational block to design \gl/\gw-free algorithms.  Finally, we note that while our algorithms provide a general purpose approach to design \gl/\gw-free algorithms, there are some open problems in this area.

\paragraph{Organization of the paper}
% \sandeepnote{Revisit}
The rest of the paper is organized as follows. In Section \ref{sec:systemmodel}, we formally define our system model.  Section \ref{sec:originaldef} recalls the definitions of wait-freedom and lock-freedom in the context of a single machine, multi-core environment. In Section \ref{sec:threadgroup}, we introduce the notion of a \threadgroup and use it to formally define the notion of \gl-freedom and \gw-freedom that extend the original definitions of lock-freedom and wait-freedom in Section \ref{sec:newdef}.
In Sections \ref{sec:singlemachine}, \ref{sec:RPC} and \ref{sec:dynamicocncurrency}, we present our \weak \gl-free implementations. 
These sections demonstrate that just as deadlock-free/lock-free/wait-free provide nuances of progress guarantees in the context of \textit{thread-based} tasks, deadlock-free/\gl-free/\gw-free provide the nuances of progress conditions in the context of \textit{\threadgroup-based} tasks. 
%, we also demonstrate the backward compatibility of \gl-freedom with lock-freedom and \gw-freedom with wait-freedom. Furthermore, we observe limitations in providing \strong versions of \gl-freedom and \gw-freedom.
%Section \ref{sec:singlemachine} presents an \sandeepnote{TODO TODO TODO }implementaiton of \weak \gl-free 
%demonstrates how the requirements of \gl-freedom and \gw-freedom can be satisfied in a system where two threads within the same system cooperate to complete the task. Section \ref{sec:RPC} extends this to multi-machine environments, and Section \ref{sec:dynamicocncurrency} extends it further to provide dynamic concurrency. 
We discuss other versions of \gl-free and \gw-free definitions in Section \ref{sec:quorum} and discuss various practical implementations and open problems in Section \ref{sec:existing}.
Subsequently, we discuss the related work and conclusion in Sections \ref{sec:related} and \ref{sec:conclusion}, respectively. 
Due to reasons of space, we discuss relaxation of various assumptions, alternative extensions, a \gl-free solution to the committee coordination problem and a blockchain application in 
the Appendix.
% \cite{ravishankar2026domainextensionlockfreedomwaitfreedomextended}.

%in Appendix \ref{sec:discuss}. We also consider alternative extensions in Appendix \ref{sec:otheralternative} that one could have considered and argue that our extension is more appropriate. Appendix \ref{sec:glfreescheduling} provides an outline for a \gl-free solution to the scheduling problem considered above. And, Appendix \ref{sec:blockchainEG} discusses another application in the context of crosschain blockchain swap. 

%Section \ref{sec:inst} demonstrates the use of the proposed definition in various types of multi-machine environments. Section \ref{sec:discuss} identifies how the assumptions made in the paper can be removed while still satisfying the requirements of lock-freedom and wait-freedom. Finally, Section \ref{sec:conclusion} presents concluding remarks. 
% \sandeepnote{Review at end.}

\section{System Model}
\label{sec:systemmodel}

The system comprises multiple threads that may be on one or more machines. 
Each machine operates under the control of its own scheduler. 
These threads can communicate with each other. We are not concerned with how the communication occurs. It could happen via messages. Or, it could happen via protocols such as RDMA that allows a machine to access memory of another machine directly. Or, it could communicate via shared file system. 
We assume communication to be eventually reliable; that is, if thread~1 attempts to communicate with thread~2 infinitely many times, then at least one of those attempts will succeed.
%\raaghavnote{\textbf{Raaghav: This paragraph already gives readers the impression (bad for us) that communication is not an atomic step, and gives the idea that this communication is also done by the same threads. } }

%For most of the paper, we will assume that only message passing based communication is available. However, we will extend it to other scenarios. 
%Each machine hosts a set of threads. 
%\footnote{Move elsewhere: Two threads on the same machine can share memory such that the operating system creates this shared memory. However, all threads can read and write from the shared memory without OS intervention. }
%
For simplicity, we assume that the number of threads (and their assignment to machines) remains fixed throughout execution: all threads are instantiated at the beginning of execution and persist indefinitely. (This is relaxed in 
the Appendix.)

The thread execution is managed by the scheduler.
%The part of the system that determines when threads take steps is called the scheduler and the order in which threads take steps is the schedule \cite{herlihybook}. 
We assume that (1) the scheduler repeatedly executes a scheduling logic to achieve this schedule, (2) the scheduler runs infinitely often, and (3) 
%every time it runs it executes a finite number of steps to complete its scheduling logic (e.g., to schedule the next task or handle one interrupt). 
% every time it runs, it completes its scheduling logic (e.g., to schedule the next task or handle one interrupt) in a finite number of steps. 
every iteration of its scheduling logic (e.g., to schedule the next task or handle one interrupt) completes in a finite number of steps. 
%We use the term \textit{\systemthread} (more on this in Section \ref{sec:RPC}) to denote the part of the system that performs these tasks. \sandeepnote{Is this still needed? } \raaghavnote{Raaghav: We otherwise need to say threads communicate through busy waiting if there is no shared memory.}
We make this scheduler requirement explicit because it streamlines modeling of interactions between cooperating threads. 

%We make this scheduler requirement explicit because it streamlines handling interrupts infinitely often. 

% \raaghavnote{The part of the system that determines when threads take steps is called the scheduler \cite{??} and the order in which threads take steps is the schedule.}
We say that a thread is in one of the two \textit{abstract} states:
\begin{enumerate*}[leftmargin=*]
    \item \executable ({intuitively, executing}): If assigned a processing unit (whether CPU, GPU, etc.), the thread will execute a step that makes progress towards completing the operation.
    \item \blocked: ({intuitively, blocked}) If assigned a processing unit, the thread cannot make progress. This could be due to waiting for a resource, waiting on a lock or busy-waiting. 
\end{enumerate*}

%\sandeepnote{CPU/GPU = Processing Unit. Instead of OS use system}} \raaghavnote{\textbf{Raaghav:} But in the second sentence of the system model, we introduce system comprising of machines, and each machine having a OS? Also do we not have one scheduler per machine? So should we say we have a number of schedulers, each of which obeys the scheduler assumptions we made earlier in this section?}

We discuss relaxation of assumptions made in the paper in
the Appendix. 
% the extended version \cite{ravishankar2026domainextensionlockfreedomwaitfreedomextended}.

\section{Wait Freedom and Lock Freedom in Single Machine Environment}
\label{sec:originaldef}
%\section{Definitions of Wait-Freedom and Lock-Freedom in a Single-Machine Environment}

In this section, we recall the traditional definitions of \emph{wait-freedom} and \emph{lock-freedom} \cite{herlihy1991wait,on-nature-of-progress}. 
%Here, all threads are \userthreads. 
Each operation is thread-based, i.e., each operation takes one thread to execute it. Each thread invokes an operation (e.g., insert, delete, lookup) on this data structure and expects that the operation will eventually return a response.
%. When a thread invokes an operation, it is expected that the operation will eventually return a response.
%
An operation is pending if a thread has invoked it but has not yet received a response.
%An operation is said to be \emph{pending} if a thread has invoked it but has not yet received a response. 
%We assume that at most one operation can be pending for any given thread at a time; that is, a thread invokes its next operation only after receiving the response to its previous one.
We assume that each thread has at most one pending operation at a time, invoking the next only after completing the previous.
%Since multiple threads may be active concurrently, multiple operations may be pending simultaneously.
A thread can execute only when the scheduler assigns it a processing unit. When a thread is assigned the processing unit, it executes one or more computational steps if it is in \executable state. If it is in \blocked state, it performs a \textit{no-op}.
%(possibly instructions for busy-waiting). 

%If the thread is \blocked (for example, waiting on a semaphore), it performs \textit{no-op} even when assigned the CPU.

\begin{definition}[Active Thread]
\label{def:activethread}
A thread is said to be \emph{active} if and only if the operating system scheduler assigns it a processing unit infinitely many times over the course of execution.
    
\end{definition}

If a thread is not active, we cannot expect it to complete its pending operation, as it may not have received sufficient steps to finish the task. With this intuition, wait-free computation is defined as follows \cite{herlihy1991wait,on-nature-of-progress}.

\begin{definition}[Wait-Freedom]
\label{def:waitfreesingle}
An operation/method $m$ is \emph{wait-free} \cite{herlihy1991wait,on-nature-of-progress} if and only if \underline{every} active thread that invokes $m$ completes it within a finite number of steps.
%if every invocation of $m$ completes  within a finite number of steps as long as the corresponding thread is active. 
%\textbf{every active thread} completes its pending operation within a finite number of its own steps.    
\end{definition}

%Note that the above definition requires an active thread, say Thread~1, to make progress without regard to actions of other threads (that may be active or not). For example, if Thread~1 is (busy-)waiting for a resource by Thread~2 and Thread~2 stops executing after a finite number of steps, Thread~1 will end up busy-waiting forever, i.e., it will not finish its method even if it executes infinitely often. In this case, the operation will not be wait-free. Likewise, if Thread~2 is active and always contending for the same resource as Thread~1 and if Thread~2 always obtains the resource then Thread~1 will fail to complete its operation even if it executes infinitely often. 

The definition of wait-freedom requires that an active thread, such as Thread~1, make progress independently of other threads. For instance, if Thread~1 is waiting on a resource held by Thread~2, and Thread~2 halts after a finite number of steps, then Thread~1 will fail to complete its operation despite executing infinitely often, thus violating wait-freedom. Similarly, if Thread~2 remains active and consistently acquires a shared resource before Thread~1, then Thread~1 may never complete its operation, again breaching wait-freedom.

%However, contention with other threads or lack of progress by other threads may prevent this.  For example, if Thread~1 is waiting for a lock held by Thread~2, and Thread~2 fails to release this lock, then Thread~1 will be unable to make progress. This situation can occur even if Thread~1 is active (assigned the CPU infinitely often) but Thread~2 is not active (assigned the CPU only finitely many times \raaghavnote{`assigned insufficiently' instead?}).

%The definitions of wait-freedom and lock-freedom aim to characterize the progress guarantees of a thread when it contends with other threads, including cases where some threads fail to make progress.

%This definition implies that if Thread~1 is active (assigned the CPU infinitely often), it will complete its operation in a finite number of steps, regardless of whether other threads—such as Thread~2—are executing, contending, or failing to make progress.

% \begin{definition}[Lock-Freedom]
% \label{def:lockfreesingle}
% An operation is \emph{lock-free} if and only if,  \textbf{at least one active thread} completes its pending operation within a finite number of steps.
% \end{definition}

\begin{definition}[Lock-Freedom]
\label{def:lockfreesingle}
An operation/method $m$ is \emph{lock-free} \cite{herlihy1991wait,on-nature-of-progress} if and only if \underline{some} active thread that invokes $m$ completes it within a finite number of steps.
%if every invocation of $m$ completes  within a finite number of steps as long as the corresponding thread is active. 
%\textbf{every active thread} completes its pending operation within a finite number of its own steps.    
\end{definition}

Note that wait-freedom and lock-freedom allow some threads to make progress even if not all threads are active, i.e., some threads stop executing. By contrast, deadlock-freedom requires that when all threads are active, some task is completed. Specifically, 
\begin{definition}[Deadlock-Freedom]
\label{def:deadlockfreesingle}
A set of tasks is \emph{deadlock-free} iff some task completes its operation when all threads are active. 
%if every invocation of $m$ completes  within a finite number of steps as long as the corresponding thread is active. 
%\textbf{every active thread} completes its pending operation within a finite number of its own steps.    
\end{definition}

From \cite{herlihy1991wait,on-nature-of-progress}, we know that wait-freedom implies lock-freedom which in turn implies deadlock-freedom. Since the definitions in these papers are based on a single-thread per operation, they cannot distinguish collaborative threads within a \threadgroup and competing threads across \threadgroups in the definition.  In fact, they classify \threadgroup based tasks to be (at most) deadlock-free. The proposed extension reintroduces the nuance between these concepts to group‑based computation.

% \spnote{From the above definitions, one can see that any algorithm that is \wf is also \lf. Further, an implementation that uses locks internally cannot be both \lf and \wf.}
% \sandeepnote{refer to MH}

To introduce the notions of \gl-freedom and \gw-freedom in this paper, we illustrate their achievability through the following two protocols: 
\begin{enumerate*}
    \item \textbf{SimpleList}. The SimpleList is a data structure that permits insert, delete and lookup. Multiple threads could invoke operations at the same time. Whenever an operation is invoked, the implementation locks the entire list, performs the operation, unlocks the list and then returns the result. It is straightforward to see that SimpleList is deadlock-free but not lock-free (hence also not wait-free).
    \item 
\textbf{Harris List \cite{Harris2001LinkedList}}. Harris List \cite{Harris2001LinkedList} implements a linked list that supports insert, delete and lookup. In this paper, we do not discuss the implementation of \cite{Harris2001LinkedList} and only rely on the fact that it is lock-free (and by implication deadlock-free).
%Observe that both SimpleList and HarrisList are also deadlock-free. 
%\spnote{There is no explanation of \df so far. Maybe this last line can be removed.}

\end{enumerate*}

\section{\ThreadGroup Semantics}
\label{sec:threadgroup}

%As discussed in Section \ref{sec:intro}, the goal of this work is to address progress conditions in tasks that are necessarily group-based, i.e., they can only be completed if a set of threads cooperate to complete a task. %Thus, the progress condition is defined over a group of threads, not individual ones.
%\footnote{To avoid the possibility of a degenerate implementation where all threads are declared to be in the same group, we require that for any thread $i$ in the group, there is some invocation of the method $m$ such that $m$ cannot complete without the execution of $i$. 
%}
%progress in a distributed system is not determined by the execution of a single thread, but rather by the coordinated execution of a group of threads. 
% SPACE For instance, in the committee coordination example from Section \ref{sec:intro}, one or more threads associated with each principal is responsible for completing the task at hand,  Execution by just a subset of these threads can never complete the task as they do not have the necessary permissions.
%For instance, in a client-server architecture, progress occurs only when both the client and server threads execute in tandem. Execution by either thread in isolation does not constitute system-level progress. 
% SPACE In other words, while the progress in a single machine environment is guaranteed when the thread executes infinitely many steps, the overall progress in distributed system will occur when each thread in the group executes infinitely many steps.
%

In this section, we start from the conventional view of a single‑threaded completed operation and generalize it naturally to group‑level tasks. Specifically, our model for each machine is identical to that in Section \ref{sec:systemmodel} and \ref{sec:originaldef}; each machine can have multiple threads, they can be in \executable or \blocked state, etc. We only add conditions that must be satisfied by a reasonable \threadgroup implementation.

%Hence, we define an operation/method/task (e.g., create common slot for a set of principals) as dependent on a set of threads forming a thread-group.
% \sandeepnote{
% \raaghavnote{In Section \ref{sec:originaldef}, }
When task completion is thread-based \cite{herlihy1991wait},
%In the traditional concurrent programs, 
the relevant threads are created before they invoke the operation and each thread is working on at most one operation at a time. We expect the same when the task is group-based, i.e., we assume that all \threadgroups are created at the beginning and each \threadgroup works on at most one task at a time. One can overcome this and allow a \threadgroup to execute multiple tasks at a time. The definitions of \gl-freedom and \gw-freedom can then be extended to permit this. In 
the Appendix
% \cite{ravishankar2026domainextensionlockfreedomwaitfreedomextended} 
we show that through modeling of some abstract threads, we can still achieve the same result with the current definition.
% }
%\spnote{The term single threaded program is misleading? Probably, single threaded program should be replaced with single machine program here and elsewhere.}

When a task completion is thread-based, the work associated with the operation is completed between the invocation and completion.
We want to ensure that the same property holds for tasks completed by a \threadgroup. Recall that threads were classified to be either in \executable state or \blocked state, where only threads in \executable state could make progress while threads in \blocked state did not do any useful work, i.e., they could be waiting/busy-waiting, etc. Since only threads in \executable state can perform actual work for an operation, we require that when the operation is invoked (respectively, completed) only the invoking (respectively completing) thread be in \executable state. (This is captured by conditions 1 and 2 in Definition \ref{def:validgroups}.)
% \spnote{Removed the repetition: We want the same property to be true for the computations performed by a \threadgroup.}

A thread is either in \executable state or \blocked state. With a \threadgroup, some threads could be in \executable state while some threads could be in \blocked state. Next, we consider transitions of a thread from \executable to/from \blocked. (1) A thread in \executable state can continue to execute its portion of the task. Once it finishes that it can transition to \blocked state. (This motivates the condition 3(a) of Definition \ref{def:validgroups}). (2) A thread in \blocked state is not doing any useful work. So, it cannot choose to transition to \executable state on its own, i.e., without interaction with some other thread in \executable state. (This captures condition 3(c) of Definition \ref{def:validgroups}.)

Finally, let's consider the interactions of multiple threads within a \threadgroup. There are various types of interactions that may happen in collaborating threads. A simple example of a client-server thread would mean that initially, the client thread is running, which activates the server thread when the operation is invoked. Upon completion, the server thread activates the client thread so that it can resume the computation further.  In a computation framework such as Hadoop \cite{Hadoop}, the initial thread may activate multiple other threads. A key requirement in this context is that the activation should occur in a finite number of steps. (This captures condition 3(b) of Definition \ref{def:validgroups}). 
% \spnote{Removed this phrase: in a finite time (i.e., }
From the above discussion, we have

\begin{definition}[Valid \ThreadGroup Conditions]
\label{def:validgroups}
For an execution of a \threadgroup $G$ associated with method $m$ to be considered valid, the following conditions must be satisfied: \begin{enumerate}[topsep=0pt, leftmargin=*]
\item \textit{Thread State at Invocation}: When any thread $T$ in \threadgroup $G$ invokes $m$  (e.g., \texttt{schedule a meeting}), all other threads in $G$ (i.e., $G-\{T\}$)
%(excluding the thread invoking the operation)
% \raaghavnote{Are we analyzing an operation as it executes or are we analyzing from its history? Other threads to be involved might not be clear in the former. If we are talking about a completed operation analyzed through a history, then the elements of the group (set) will be well defined to describe a condition for `other theads'.}
must be in the \blocked state. $T$ is in \executable state. 

\item \textit{Thread State upon Completion}: When $m$ is completed, all threads in $G$ (except for the thread invoking it) are in \blocked state. The thread invoking the operation is in \executable state. 
%\footnote{We could permit some threads in the \threadgroup to continue doing some asynchronous task. It is discussed in Appendix \ref{sec:discuss}.}
%It is not required that all threads become \blocked when the operation completes. This could happen if some task is being done asynchronously with the method call. However, it is required that this asynchronous work eventually finishes.

\item \textit{State Transitions}: 

\begin{enumerate} 
\item A thread in  \executable state may transition to the \blocked state at any time. 

\item A thread, say $T_i$, in \executable state can wake up a thread, say $T_j$, in \blocked state, causing $T_j$ to transition to \executable state in a finite number of steps.
%Furthermore, this is achieved by executing only a finite number of steps in thread $i$,  $j$ and possibly some other \textit{\systemthreads} that assist execution of \threadgroups. 
%\raaghavnote{We need an example with a completely analyzed execution (including any potential synchronization points that are avoided in the implementation of the example) that is measured to involve only a finite number of steps under contention to justify this. This is the make or break portion for this work.}
%\sandeepnote{Added how to implement RPC in Section 6}

\item A thread in \blocked state cannot transition to \executable state independently; it requires an interaction from another thread in \executable state. 

\end{enumerate} 

\end{enumerate}

\end{definition}

This paper limits definitions of \gl-freedom and \gw-freedom to valid \threadgroup computations. Applying it in other contexts is feasible with some abstract modeling. We discuss this in 
the Appendix.

\section{Wait Freedom and Lock Freedom in \ThreadGroups}
\label{sec:newdef}

%\section{Extension of Wait-Freedom and Lock-Freedom to Distributed Systems}

% SPACE In this section, we extend the definitions of \emph{wait-freedom} and \emph{lock-freedom} to group computations and demonstrate that they have the same semantic interpretation of \cite{herlihy1991wait,on-nature-of-progress} for single-machine systems.
%\raaghavnote{\textbf{Raaghav: Maybe swap weak and strong active? Also the ``equivalently" sentence seems unnecessary here and better off directly in the gl, gw freedom definitions}}

We begin by extending the definition of an active thread (Definition \ref{def:activethread}) and utilize it to extend the definition of lock-freedom and wait-freedom. 

\begin{definition}[\WeakActive \ThreadGroup]
\label{def:weakactivethreadgroup}
A \threadgroup is said to be \emph{\weakactive} if and only if \textbf{all}
%\footnote{We could replace the word `all' with `sufficiently many'. For example, if there are two threads A1 and A2 to perform task for node A and B1 and B2 to perform task for node B. However, the task can still be done if one thread on A and one thread on B is active. 
% However, this assumes knowledge about internals of the \threadgroup. Hence, it is outside the scope of the paper.
%} 
threads within that group are active. Equivalently, a \threadgroup is \weakactive precisely when each of its constituent threads is scheduled to execute infinitely often.
\end{definition}

\begin{definition}[\StrongActive \ThreadGroup]
\label{def:strongactivethreadgroup}
A \threadgroup is said to be \emph{\strongactive} if and only if \textbf{some}
threads within that group are active. Equivalently, a \threadgroup is \strongactive precisely when some of its constituent threads is scheduled to execute infinitely often.
\end{definition}

The above two definitions are at the extreme of what an active thread group can be. Definition \ref{def:weakactivethreadgroup} requires all threads in a group to be active while Definition \ref{def:strongactivethreadgroup} requires at least one thread to be active. One can consider intermediate variations where some subset of threads are active. We will discuss this in Section \ref{sec:quorum}.

\iffalse
\noindent\textbf{Justification for Definition \ref{def:activethreadgroup}. }
As discussed in Section \ref{sec:intro}, the completion of an operation requires the cooperation of multiple threads because no thread can do all the work. %This cooperation is represented by the concept of a \emph{\threadgroup}. 
%Since threads in a \threadgroup collaborate, they all must execute to complete the operation. 
In other words, if any thread within a \threadgroup fails to execute, it is possible that the operation will not complete.
Expecting that a \threadgroup makes progress even if one of its members fails is equivalent to expecting that A1 and B1 schedule a meeting time even though B1 does not execute and calendar of B is only available to B1. 
\fi

Just like lock-freedom and wait-freedom where only active threads could complete their task, we expect only active groups to be able to complete their task. With this motivation, we now define the notion of \gl-freedom and \gw-freedom. Note that these definitions can be instantiated with \strongactive or \weakactive \threadgroups. 

%As noted in Section \ref{sec:originaldef}, progress was only expected from active threads. 
%f a thread, say $T_1$, of group $G_1$ is inactive, i.e., it executes only a finite number of times, $T_1$ may not finish the work expected of it in $G_1$. Since this work cannot be completed by others, the operation of $G_1$ may remain incomplete. 
%If a thread is inactive, i.e., it executes only a finite number of times, the CPU allocation it receives may be insufficient to complete its assigned operation. Similarly, if a thread within a group, say $T_1$, does not receive adequate CPU time, it may fail to fulfill its portion of the task. Since other threads in the group cannot substitute for $T_1$, the overall operation remains incomplete. Therefore, progress can only be expected from \threadgroups in which every thread has had ample opportunity to execute.

 Based on this discussion and the notion that an operation is wait-free if every operation finishes in a finite number of steps, we are now ready to extend the definition of wait-freedom to group environment. 

% SPACE Specifically, since a \threadgroup cannot complete its task if insufficient CPU is assigned to it, wait-freedom is defined as follows 
%(As stated in System Model, these definitions rely on the assumption that \systemthreads execute infinitely often):

%\subsection{Distributed-System Progress Conditions}

%We extend the progress conditions to distributed systems as follows:

% \begin{definition}[\gw-freedom]
% \label{def:waitfreedomdistributed}
% Let $G=\{G_1, G_2, ...\}$ be a set of pairwise disjoint groups  where each group is a set of threads. 
% An operation is \emph{wait-free} for $G$ if and only if,  for any $i$, if $G_i$ is active then $G_i$ completes its task 
% %\textbf{every active \threadgroup} completes its pending operation 
% within a finite number of its  own steps. %\sandeepnote{(under the assumption that every \systemthread is  active)}
% \end{definition}

\begin{definition}[\Strong/\Weak\gw-Freedom]
\label{def:waitfreedomdistributed}
An operation/method $m$ is \emph{\strong \gw-free} (respectively, \weak \gw-free) if and only if every \strongactive (respectively, \weakactive) \underline{\threadgroup} that invokes $m$ completes it within a finite number of steps.
%if every invocation of $m$ completes  within a finite number of steps as long as the corresponding thread is active. 
%\textbf{every active thread} completes its pending operation within a finite number of its own steps.    
\end{definition}

% \begin{definition}[wait-freedom]
% \label{def:waitfreedomdistributed}
% Let $T=\{T_1, T_2, ...\}$ be a set of threads. 
% An operation is \emph{wait-free} for $T$ if and only if,  for any $i$, if $T_i$ is active then $T_i$ completes its task 
% %\textbf{every active \threadgroup} completes its pending operation 
% within a finite number of its own steps. %\sandeepnote{(under the assumption that every \systemthread is  active)}
% \end{definition}

\begin{definition}[\Strong/\Weak\gl-freedom]
\label{def:lockfreedistributed}
An operation/method $m$ is \emph{\strong \gl-free} (respectively, \weak \gl-free) if and only if some \strongactive (respectively, \weakactive) \underline{\threadgroup} that invokes $m$ completes it within a finite number of steps. 
%\sandeepnote{(under the assumption that every \systemthread is  active)}

\end{definition}

%\spnote{There is some confusion in the definitions}

% \begin{definition}[\gl-freedom]
% \label{def:lockfreedistributed}
% An operation is \emph{lock-free} if and only if,  \textbf{at least one active \threadgroup} completes its pending operation within a finite number of their own steps.
% %\sandeepnote{(under the assumption that every \systemthread is  active)}

% \end{definition}

%\sandeepnote{add to concl}
%\textbf{Semantic equivalence to original definitions \cite{herlihy1991wait}}

\noindent\textbf{Implications of Definition \ref{def:waitfreedomdistributed} and \ref{def:lockfreedistributed} in the context of \strong \gl-freedom and \strong \gw-freedom. }
In the context of \strong \gl-freedom and \gw-freedom, task assigned to $G$ will complete if at least one of the threads in $G$ makes progress. In other words, even dependency between threads within a \threadgroup is not permitted. 

\noindent\textbf{Implications of Definition \ref{def:waitfreedomdistributed} and \ref{def:lockfreedistributed} in the context of \weak \gl-freedom and \weak \gw-freedom. }
In the context of \weak \gl-freedom and \gw-freedom, Definition \ref{def:waitfreedomdistributed} requires that every active \threadgroup must complete its operation. If all threads in a group (e.g., $G_1$ with $T_1$ and $T_2$) execute infinitely often, the group must make progress.
%We observe that the preceding definition mandates that every active \threadgroup must successfully complete its operation. That is, if all threads within a given group are scheduled to execute infinitely often, the group as a whole is required to make progress. Consider a \threadgroup $G_1$ comprising two threads, $T_1$ and $T_2$. 
If $T_1$ is blocked awaiting progress from $T_2$, this poses no issue: since $T_2$ is scheduled infinitely often, it will eventually complete its task, thereby enabling $T_1$ to proceed. Crucially, however, the definition stipulates that $G_1$ must make progress independently of the behavior of any other \threadgroup (say $G_1'$). For instance, if a thread, say $T_1'$ in $G_1'$ (consisting of $T_1'$ and $T_2'$)  halts prematurely---rendering both $T_1'$ and $G_1'$ inactive---this must not impede the completion of $G_1$'s operation. 

\subsection{Backward Compatibility with Original Definition from \cite{on-nature-of-progress, herlihy1991wait}}
\label{sec:backward}

%\spnote{Should this short content be treated as a separate section?}

Before we discuss the use of the Definitions  \ref{def:waitfreedomdistributed} and \ref{def:lockfreedistributed}, we note that these definitions are extensions  of the original definitions. Specifically, if the \threadgroup consists of just one thread, then an active \threadgroup (assigning infinite steps to each thread) is equivalent to an active thread (assigning infinite steps to that thread). Thus, we have:

\begin{observation}
    If each \threadgroup consists of a single thread, then both Definitions \ref{def:weakactivethreadgroup} and  \ref{def:strongactivethreadgroup} are the same as Definition \ref{def:activethread}. In turn, Definition \ref{def:waitfreedomdistributed} (respectively Definition \ref{def:lockfreedistributed}) is identical to Definition \ref{def:waitfreesingle} (respectively, Definition \ref{def:lockfreesingle}).
 \end{observation}
 %\raaghavnote{\textbf{Raaghav: What is a group problem?}}

 Note that the above observation is independent of whether one is using the \strong version or the \weak version; when a \threadgroup consists of only 1 thread, these two definitions are identical. 

\subsection{Impossibility Result for \Strong \gl-freedom/gw-freedom for Non-Trivial Group Problems}

Although Definitions \ref{def:weakactivethreadgroup} and \ref{def:strongactivethreadgroup} are compatible with the original definition, they differ substantially by the \threadgroups comprising more than one thread. Definition \ref{def:strongactivethreadgroup} is closer in spirit to the original definitions\cite{on-nature-of-progress}. For example, a thread completes its task in a wait-free manner if it executes infinitely often. Definition \ref{def:strongactivethreadgroup} requires the same even if the \threadgroup consists of multiple threads. By contrast, Definition \ref{def:weakactivethreadgroup} requires all threads in a \threadgroup to be active for that \threadgroup to complete the task. Unfortunately, \strong \gl/\gw-freedom is not feasible in most contexts.

To formalize this, we say that a task is group-based if there is a group of threads assigned to complete it. We say that group-based task is non-trivial if execution of at least two threads is necessary to complete that task. 

An example of a group-based non-trivial task would be a client server application where at least one thread of the client and one thread of the server is required to execute. Another example would be a partitioned data structure where at least one thread must execute in each partition to complete the task. A task such as multi-threaded sorting or multi-threaded search is not non-trivial, as one thread can complete the task and multiple threads are only useful for performance. Based on this idea, we define
%For such non-trivial tasks, we have

\begin{definition}\label{def:nontrivial}
    We say that a \textit{group-based} task is non-trivial if completing that task \textbf{requires} execution by at least two threads.
\end{definition}

We note that non-triviality does not require that a specific set of two threads have to execute. A task would be non-trivial even if it is necessary that two threads out of $\{T_1, T_2, T_3\}$ execute. 
% \sandeepnote{Drop if needed}

\begin{theorem}
    There is no algorithm that provides \strong \gl-freedom or \strong \gw-freedom for a non-trivial task. 
    \label{thm:nostrong}
\end{theorem}

% \begin{proof}
%     %This result follows from the Definition \ref{def:strongactivethreadgroup} and \ref{def:nontrivial}. 
% \newtext{
%     Let $T$ be a non-trivial task with \threadgroup $G$. 
%     Definition \ref{def:waitfreedomdistributed} (respectively, Definition \ref{def:lockfreedistributed}) requires that a \strong \gw-free (respectively, \strong \gl-free) algorithm completes $T$ even if $G$ is \strongactive. To complete the proof, we show that even if $G$ is \strongactive, $T$ may fail to be completed. 
%     %We show that if $G$ is $\strongactive$ , $T$ may still fail to be completed. 
%     %We show this result by demonstrating that $T$ cannot be completed with \strong \gl-freedom (respectively, \gw\-freedom).
%     In the context of \strong \gl/\gw-freedom, by Definition \ref{def:strongactivethreadgroup}, it suffices if only one thread in $G$ to be active to ensure that $G$ is \strongactive.    
%     By Definition \ref{def:nontrivial}, $T$ requires execution by at least two threads, i.e., task of $T$ cannot be completed if only one thread is active. 
    
%    % If this thread executes alone, by definition $T$ being non-trivial, $T$ cannot be completed. 
% }
% \end{proof}

\begin{proof}
    
We prove this by proof-by-contradiction. Assume that there is an Algorithm $\mathcal{A}$ that provides \strong \gl-freedom (or \gw-freedom). 
Consider the execution of $\mathcal{A}$ where there is only one non-trivial task $T$ running on \threadgroup $G$.     Definition \ref{def:waitfreedomdistributed} (respectively, Definition \ref{def:lockfreedistributed}) requires that $\mathcal{A}$ completes $T$ even if $G$ is \strongactive, i.e., only one thread in $G$ is active. Since $T$ is a non-trivial task, by Definition \ref{def:nontrivial}, execution of at least two threads is needed to complete $T$. Hence, $T$ will not be completed thereby contradicting that $\mathcal{A}$ is a \strong \gl-free (or \gw-free) algorithm. 
\end{proof}

\textbf{Based on this, unless otherwise stated, in the subsequent sections, \gl-freedom/\gw-freedom refers to \weak \gl-freedom/weak \gw-freedom.}
\section{Implementation of \weak \gl-freedom for protected linked list: Under Identical Assumptions in \cite{herlihy1991wait,on-nature-of-progress} }
\label{sec:singlemachine}

Since \strong \gl-freedom and \gw-freedom is impossible for non-trivial group problems, in this section, we provide an implementation of \gl-freedom for a \textit{protected linked list} data structure which also forms a foundational block for further analysis in Sections \ref{sec:RPC} and \ref{sec:dynamicocncurrency}. %\spnote{Are \gl-freedom and \gw-freedom the same as weak \gl-freedom and \gw-freedom? If so, we must mention this equivalence clearly. }

% \newtext{
%The problem is as follows: There are a set of threads, $T_1, T_2,\cdots$ that invoke operations on the list that are completed with the assistance of threads $T_1', T_2, \cdots$. One example for this problem is that the list is managed by a process, say $L$ that allows other processes to access the list. But the list can only be accessed by a thread of $L$. In this case, $T_1', T_2', \cdots$ will be threads of $L$ while $T_1, T_2, \cdots$ belong to other processes that want to use the list $L$. 

    The setting is as follows: We have a collection of independent threads $T_1, T_2, \cdots$ that request operations on a shared list. These operations are not executed directly by the requesting threads; instead, they are carried out by a corresponding set of helper threads  $T_1', T_2', \cdots$. A representative example is a system in which the list is owned and managed by a dedicated process $L$. External processes may submit requests to manipulate the list, but only threads belonging to $L$ are permitted to access or modify it. In such a scenario, the helper threads $T_1', T_2'\cdots$ are the internal worker threads of process $L$, while the requesting threads $T_1, T_2, \cdots$ belong to other processes that rely on $L$  to perform list operations on their behalf.
    This is similar to the case where a process updates kernel data structures via system calls that are routed through interprocess communication. However, in this case, the data structure is to be implemented entirely in user space. 
    Just as in kernel data structures, there may be some checks that $L$ wants to do (e.g., inserted items are within acceptable range) and, hence, we do not want $T_1$ to directly access the list. In other words, when $T_i$ starts the operation, $T_i'$ validates it and, if valid, performs the operation on the list. Otherwise, it returns an error and makes no changes to the list.
    % }
%\spnote{The justification of this example is not clear.}

Observe that there is no lock-free implementation of protected linked list, as the thread that can invoke the operation cannot complete the operation and vice versa. We show that HarrisList can be extended to \gl-free protected linked list. If one applies the same extension to SimpleList then it is not \gl-free. Note that the same idea could be applied to any lock-free/wait-free data structure, our extension is independent of the actual HarrisList implementation.

%Here, the data structure, say a list, is managed by threads $T_1', T_2'\cdots$ using an algorithm such as Harris Linked List \cite{Harris2001LinkedList}. The data structure is accessed by client threads $T_1, T_2, \cdots$.  
%The protection layer can enforce any collection of constraints, such as restricting insertions to items with a prescribed structure or ensuring per‑thread monotonicity of inserted values. In our setting, the essential requirement is that $T_1'$ validates these conditions and, upon success, invokes the task. If a violation is detected, $T_1'$ performs no operation and returns execution to $T_1$.
%Clearly, no lock-free solution exists for such protected data structure, as two threads (one in client space and one in protected space) must execute to complete the task at hand.  

\textbf{Outline of Algorithm to implement Protected Linked List: }
At the beginning of the program (i.e., independent of the operations being invoked), we create a shared memory for $T_1$ and $T_1'$ (using primitives such as shmget/shmat \cite{shmget_manpage}). 
%We assume that there is some limited shared memory available to both of them. 
While the OS is responsible for creating this shared memory, during execution, the threads can access it without OS intervention, i.e., a system call is not needed when accessing this memory. This is important since requiring a system call would make it impossible to meet the constraints of Definition \ref{def:validgroups} as there is no control over the number of steps that may be needed during this system call. In our implementation, we ensure that the only operations permitted on this shared memory are read and append.
%This example is an adaptation of the RPC model from the previous subsection, but with \textbf{C1} and \textbf{S1} on the same machine, communicating via shared memory. 
%
To execute an operation, say insert in a protected linked list, the actions are as follows:\footnote{We note that this implementation as well as the one in next two sections uses several ideas from ZeroMQ \cite{zeromq}. Details of this are discussed in Section \ref{sec:existing}.
}
%. We discuss this in detail in Section \ref{?} to demonstrate that this implementation can be achieved in practice with very low overhead. We note that the proposed implementation can be achieved in a straightforward manner with ZeroMQ and RDMA thereby demonstrating its practicality.}

\begin{itemize}[leftmargin=*]
    \item First, $T_1$ writes the details of the operation in shared memory. This is achieved via writing an appropriate structured record with appropriate delimiters that allows $T_1'$ to know when the record is complete. Afterwards $T_1$ busywaits on the shared buffer until $T_1'$ provides the result. 
    \item Initially (and between two consecutive operations), $T_1'$ repeatedly (i.e., infinitely often) reads the shared memory until it finds that $T_1$ has written the complete record of requested operation.
    Then, it does the necessary validation checks and updates the protected data structure. Then, it writes this result into shared memory and returns to reading the shared buffer to find the next invocation. 
    \item $T_1$ repeatedly reads the shared memory. When $T_1$ notices that the data is written to the shared memory, $T_1$ returns that result as the result of the invoked operation. 
\end{itemize}

\begin{theorem}
    The above implementation  satisfies the constraints of Definition \ref{def:validgroups} if the thread groups are $\{ \{T_1, T_1'\}, \{T_2, T_2' \} \cdots \}$.
    %and $\{T_1', T_2', \cdots\}$ implement Harris List \cite{Harris2001LinkedList}.
\end{theorem}

\begin{proof}
For thread group $\{T_i, T_i'\}$, only $T_i$ invokes the operation. When that happens, by design $T_i'$ is in \blocked state. When $T_i$ completes that operation, $T_i'$ is again in \blocked state. Thus, the first two conditions are satisfied. 

Condition 3(a) and 3(c) are trivially satisfied. Regarding condition 3(b), writing method parameters to shared memory takes only a finite number of steps of $T_i$, as $T_i$ is the only thread writing to the shared memory of $\{T_i, T_i'\}$ at that time. After a finite number of steps of $T_i'$, $T_i'$ will notice the completed structure and invoke the operation. In other words, when $T_i$ wants to activate $T_i'$ it is achieved by finite steps of $T_i$ and $T_i'$. Using the same argument, the property holds when $T_i'$ completes the operation and returns the result to $T_i$. 
\end{proof}

Note that the above result is independent of whether $\{T_1', T_2', \cdots\}$ implement HarrisList or SimpleList. However, using the former makes it a \gl-free list while using the latter it does not. Specifically, 

\begin{theorem}
    The above implementation is \weak \gl-free if $\{T_1', T_2', \cdots \}$ implements HarrisList \cite{Harris2001LinkedList} where the groups are $\{\{T_1, T_1'\}, \{T_2, T_2'\}, \cdots \}$.
    \label{thm:harris-single-glfree}
\end{theorem}

\begin{proof}
Suppose $S$ is the set of \weakactive \threadgroups. First, we consider the behavior of threads in $S$. Let $\{T_i, T_i'\} \in S$. Since both $T_i$, $T_i'$ execute infinitely often then $T_i$ will eventually write the request to shared buffer, $T_i'$ will read that request and $T_i'$ will execute the algorithm in \cite{Harris2001LinkedList} to complete that operation. 

Second, we consider behavior of threads outside $S$. Let $\{T_k, T_k'\} \not \in S$. Hence, either $T_k$ or $T_k'$ does not execute infinitely often. 
If $T_k'$ does not execute infinitely often, it cannot prevent threads in $S$ from making progress after it has stopped executing. If $T_k'$ executes infinitely often thereby implying that $T_k$ does not execute infinitely often, $T_k'$ cannot get new operations invoked by $T_k$. In other words, $T_k'$ executes infinitely often to either complete the last task invoked by $T_k$ but never completes the task, or simply stay in \blocked state waiting for an operation.
In all three cases for $T_k'$ (\textit{viz.} finite steps, continuous failing of the last operation and staying in BL state), $T_k'$ does not affect the progress of the thread groups in $S$.
% Summarizing this, if $\{T_k, T_k'\} \not \in S$ then either $T_k'$ does not perform its last operation successfully or $T_k$ does not assign infinitely many operations to $T_k'$. 

Let $H$ be the list of active threads from the set $\{T_1', T_2', \cdots\}$ and consider the suffix of the execution history \cite{on-nature-of-progress, linearizability} where all non-active threads have stopped executing. By property of HarrisList \cite{Harris2001LinkedList}, one of the threads in $H$ will complete the operation. From the above discussion, if $\{T_i, T_i' \} \in S$ then $T_i' \in H$. Also, if $\{T_i, T_i' \} \not \in S$ then either $T_i'$ does not complete its last operation in this suffix or $T_i'$ does not have any task to execute in that suffix. It follows that there is some $T_j'$ that completes its operation such that $T_j' \in H$ and $\{ T_j, T_j' \} \in S$. 

%, by \cite{Harris2001LinkedList}, some thread $T_j'$, where $\{T_j, T_j'\} \in S$, completes the list operation in protected space.

Since $\{T_j, T_j'\} \in S$, $T_j$ also executes infinitely often. Since the interaction between $T_j$ and $T_j'$ is not dependent on other threads, $T_j$ will eventually complete the operation it invoked. 

\end{proof}

\begin{theorem}
    The above implementation is not \gl-free if $\{T_1', T_2', \cdots \}$ implements SimpleList where the groups are $\{\{T_1, T_1'\}, \{T_2, T_2'\}, \cdots \}$.
    \label{thm:simplelistnotglfree}
\end{theorem}

\begin{proof}
    Consider the case $T_1$ and $T_2$ invoke operation and activate $T_1'$ and $T_2'$. Now, $T_1'$ locks the list but does not execute further. Then $T_2'$ cannot complete its task. Thus, even if $\{T_2, T_2'\}$ is an active thread group, it is not able to complete its operation. Hence, the above implementation is not \gl-free if $\{T_1', T_2', \cdots \}$ implements SimpleList where the groups are $\{\{T_1, T_1'\}, \{T_2, T_2'\}, \cdots \}$. 
\end{proof}

\begin{theorem}
    Given a system with threads $\{ T_1, T_1', T_2, T_2', \cdots \}$, the above implementation is deadlock-free but not lock-free whether one uses HarrisList or SimpleList. 
\end{theorem}
\begin{proof}
    If all threads execute infinitely often, at least one operation completes. Hence, deadlock-freedom follows. Lack of lock-freedom follows from the fact that if $T_1$ executes alone, no operation is completed (since no steps are scheduled for any $T_i'$). 
\end{proof}

\noindent{\textbf{Revisiting Why Extend the Definition of lock-freedom/wait-freedom}}
At this point, we recall that both HarrisList and SimpleList are deadlock-free. HarrisList is lock-free and SimpleList is not lock-free.  Under the definitions from \cite{herlihy1991wait, on-nature-of-progress}, extensions of these to protected list are only deadlock-free. But neither of these extensions are lock-free as defined by \cite{on-nature-of-progress, herlihy1991wait}. In other words, the nuance of lock-freedom is lost. Our definition of \gl-freedom allows one to provide this nuance. And, this is the main purpose of extending the definition to \threadgroups.

\section{\gl-free List in Multi-Machine Environment}
\label{sec:RPC}
In this section, we present \weak \gl-free list in a multi-machine (client-server) setting. This extends the implementation of protected list from Section \ref{sec:singlemachine}. It is straightforward that there is no lock-free implementation of multi-machine linked list as a server and a client thread would be needed to complete the operation. We show how HarrisList can be extended to \gl-free list while the same extension fails to be \gl-free for SimpleList. 
%\spnote{The above paragraph needs to be corrected}

Our implementation makes the following assumptions: 
\begin{enumerate*}
    \item Sending a message involves a finite number of steps. After these finite steps, either the message is sent successfully or the message is lost in transit. 
    %on $C1$ to send a message to $S$, either the NIC thread on $S$ receives the message or the message is lost in transit. This is essentially equivalent to a timeout mechanism on machine $C1$. It follows that it is satisfied by existing systems. 
    \item If a message is sent infinitely often by machine $i$ to $j$, it is eventually received at node $j$. Observe that without these assumptions, it is impossible to build a valid \threadgroup that satisfies Condition 3(b) across multiple machines. At the end of this section, we show how this implementation is compatible with the implementation in the current version of Linux. In Section \ref{sec:existing}, we also provide an alternate implementation that uses ZeroMQ \cite{zeromq} and RDMA without relying on the kernel.
\end{enumerate*}

% Here, we demonstrate the definition of \gl-freedom and \gw-freedom in a client-server model where the client invokes the operation using RPC (or similar mechanism) for a data structure held by the server. We note that the current implementation of RPC in Linux does not meet the requirements of \gl-freedom and \gw-freedom but our proposed implementation does. 

% Consider two implementations of the linked-list in a client-server setting where the server maintains the list and the clients invoke operations such as insert/delete/lookup. The first implementation locks the entire list while the second uses a wait-free list (e.g. \cite{timnat}). 
% Clearly, the former will not satisfy the Definition \ref{def:waitfreedomdistributed}. We show that the latter satisfies \gw-freedom requirements of Definition \ref{def:waitfreedomdistributed} if RPC is implemented using the approach in this section. %(We also demonstrate that the current Linux implementation of RPC fails to satisfy \gw-freedom). 

\textbf{Outline of Algorithm to implement Multi-machine Linked List: }
Initially, for simplicity, we assume the existence of threads that manage the network interface card (NIC).
%threads that make assumptions similar to that of scheduler from Section \ref{sec:systemmodel} to handle the tasks of network interface card (NIC). 
Then, we show how these additional \systemthreads can be removed without losing the \gl-free property by updating the scheduler that {resembles} the current scheduler in the Linux operating system. 

\textit{Thread Groups. } For each client $C_1$, there is a \threadgroup $\{ T_{C_1}, T_{S_1} \}$ where $T_{C_1}$ runs on $C_1$ and $T_{S_1}$ runs on server. There could be multiple clients on the same physical machine. Additionally, there are 2 NIC threads per machine, one for send and one for receive. All clients on the machine (respectively, all threads on the server) share these NICs. And, as stated above, we will remove the need of these threads later. 

%Clearly, to demonstrate that an implementation of the client-server system satisfies \gw-freedom, we will need additional \systemthreads that handle network communication. Initially, we assume the existence of threads that are similar to the scheduler, i.e., we assume (1) that they execute infinitely often, and (2) when they execute, they complete in finite steps (even if that means that the operation is not completed). We show how this assumption can be removed by updating the scheduler. We also note that our proposed approach to update the scheduler {resembles} the current Linux operating system. 

\textit{Invocation of List Operation. }
Similar to Section \ref{sec:singlemachine}, shared memory is established between $T_{C_1}$ and the \textit{sender NIC thread}. When $C_1$ invokes a list operation, $T_{C_1}$ writes the record of the request in this shared buffer. The sender NIC thread  busy waits on all the buffers shared with it in a round-robin manner. When it identifies a complete record of the request, it begins the process of sending this request to the server. Since the send may fail, this NIC thread will continue to send all pending requests on that machine until a response to that request is received. A sequence number (shared amongst all $T_{C_i}$) is added to uniquely identify each request. 

Since the sender NIC thread at $C_1$ attempts to send any pending request infinitely often, the NIC thread on $S$ eventually receives this message. After receiving the message, it writes it into the memory shared with $T_{S_1}$. 
%a wait-free queue (\cite{??}). This guarantees that the added task in receiving the message is finite thereby preserving the assumption about the number of steps involved in receive.  The second NIC receiver thread reads from this shared queue. This thread shares a separate memory with each server thread. Upon de-queuing the request from the queue, it adds that request in the shared memory with $T_{S_1}$. 
Upon finding a new operation in its shared memory, $T_{S_1}$ performs that operation on the list. 

Returning the result to the client is performed in the same fashion, where $T_{S_1}$ writes the result to be read by the sender NIC thread at the server, which transmits it back to the client which writes it into the shared memory of $T_{C_1}$. 
%This action also tells the sender NIC thread at $C_1$ that the operation was completed and, hence, there is no need to send it again. 

\textit{Removing the need for NIC threads for managing network. } \
We assumed the existence of NIC threads to manage network traffic. This assumption can be removed. All we assume about the NIC threads is that (1) they execute infinitely often and (2) when they run, they complete in finite steps (and either send the message or the message is lost). The scheduler also satisfies these properties. Hence, we can remove the NIC threads if we require that the scheduler performs the task of NIC threads. (Scheduler does not need to do this at every invocation; just doing it infinitely often suffices.) Specifically, $T_{C_i}$ and $T_{S_i}$ will share memory with the scheduler on their respective machine. When the scheduler is invoked, it checks if there is any pending message that needs to be transmitted and then transmits it. And,  it does that work before scheduling another thread. With this stipulation, there will be no need for the NIC threads. One may wonder if adding this responsibility to the scheduler is reasonable. In fact, the current Linux implementation mimics this. Specifically, networking is handled by DMA, interrupt handler (top-half) and SoftIRQ (bottom-half). Generally, all of these have higher priority than kernel threads, i.e., they get scheduled before any user/kernel threads. In other words, they are scheduled (even if not explicitly by the system scheduler) before all user/kernel tasks. This is equivalent to the scheduler doing these jobs by itself but completing it before scheduling any other threads. 

\begin{theorem}
    The above implementation is \weak \gl-free if $\{T_{S_1}, T_{S_2}, \cdots \}$ implement Harris List \cite{Harris2001LinkedList} where the groups are $\{\{T_{C_1}, T_{S_1}\}, \{T_{C_2}, T_{S_2}\}, \cdots \}$.
\end{theorem}

\begin{proof}
Suppose that $\mathcal{A}$ is the set of \weakactive \threadgroups. 
Let $\{ T_{C_i}, T_{S_i} \}$ be one of the \threadgroups in $S$.
Since $\{T_{C_i}, T_{S_i}\}$ is \weakactive, i.e., both $T_{C_i}$, $T_{S_i}$ execute infinitely often then $T_{C_i}$ will eventually write the request to shared buffer (with its own scheduler). 
Since the scheduler where $C_i$ is located runs infinitely often and checks for all pending requests in a round-robin manner, it will attempt to send this request. Since all pending requests are transmitted infinitely often, this request will be received by the scheduler on $S$. This scheduler will now write this request in the memory it shares with $T_{S_i}$. Since  $\{T_{C_i}, T_{S_i}\}$ is \weakactive, $T_{S_i}$ will eventually read this request and execute the corresponding request, i.e., it  will execute the algorithm in \cite{Harris2001LinkedList} to implement that operation. Thus, if $\{ T_{C_i}, T_{S_i} \} \in \mathcal{A}$ then $T_{S_i}$ will be active thread, i.e., execute infinitely often. 

Similar to the analysis of Proof of Theorem \ref{thm:harris-single-glfree}, other threads on the server may be active, i.e., execute infinitely often. But that will happen only if those threads never complete the last task assigned to them.

Using the above two results, by \cite{Harris2001LinkedList}, some thread $T_{S_j}$, where $\{T_{C_j}, T_{S_j}\} \in \mathcal{A}$, completes the list operation in protected space. Since  $\{T_{C_j}, T_{S_j}\}$ is \weakactive,  $T_{C_j}$ also executes infinitely often. Since the interaction between $T_{C_j}$ and $T_{S_j}$ is not dependent on other threads, $T_{C_j}$ will eventually complete the operation it invoked. 
\end{proof}

\begin{theorem}
    The above implementation is not \weak \gl-free if $\{T_{S_1}, T_{S_2}, \cdots \}$ implement SimpleList  where the groups are $\{\{T_{C_1}, T_{S_1}\}, \{T_{C_2}, T_{S_2}\}, \cdots \}$.
\end{theorem}

\begin{proof}
    The proof of this is similar to Theorem \ref{thm:simplelistnotglfree}
\end{proof}

\section{\gl-free Distributed Partitioned List }
\label{sec:dynamicocncurrency}
In this section, we show how to extend the algorithm in Section \ref{sec:RPC} to \weak \gl-free implementation of distributed partitioned list. The goal of this section is to demonstrate that our definitions of \gl-freedom and \gw-freedom are more generic than those where the execution is sequential.

We consider an extension of HarrisList to HarrisListUnsorted. Specifically, in this case, insert specifies the place (pointer and server identity to previous key) where the key is to be inserted. Hence, the list itself may be unsorted. And an element may be present more than once. We consider a similar extension for SimpleListUnsorted. For the sake of simplicity, we assume that the list is partitioned and static. We note that dynamically distributed list is considered in \cite{DBLP:journals/corr/abs-2510-06387}. The same analysis can be used to demonstrate that \cite{DBLP:journals/corr/abs-2510-06387} is \weak \gl-free, but focusing on it is outside the scope of the paper. 

HarrisListUnsorted and SimpleListUnsorted support Lookup, Insert and Delete. For the sake of simplicity, we assume that the semantics of Lookup(key) is that it returns true if the key was continuously present. It returns false if the key was never present. However, it may return true or false if the key was present only for a part of the time.

\textbf{Outline of algorithm to implement Partitioned Linked List: }
We present our approach with two NIC threads per machine. As shown in Section \ref{sec:RPC}, we can obtain a \gl-free version of the list that removes these NIC threads. 

\textit{Thread Groups.}
For each client $C_i$, there is a \threadgroup $\{ T_{C_i}, T_{S_1}^i, T_{S_2}^1, \cdots T_{S_k}^1\}$ where $S_1$, $S_2 \cdots S_k$ are servers that host the list. Each server has a thread dedicated to each client. This thread would be used for operations invoked by that client. 

The operations invoked by the client remain similar to that in Section \ref{sec:RPC} with a few subtle differences. A Lookup operation would need to be sent to each server. Hence, the client will share multiple buffers with its sender NIC thread, one for each server. Note that in Section \ref{sec:RPC}, the client thread went into \blocked state when it wrote the record for transmission of the request. Instead, it would do so after writing all the records. Likewise, it will wait for all the responses before changing from \blocked state to \executable state. Observe that in this case, the operation begins with a single thread on the client that is in \executable state. In the middle of the operation, there are multiple threads, possibly one on each server, that is in \executable state. 

For insert/delete operation, only one server would be involved, as the operation parameters already identify where the item is to be inserted. In this case, the client will be sending only one message. In other words, the threads on the other servers will play no role in the operation. However, since insert operation may require their execution in some instances (e.g., when the operation is being performed on that machine), they need to be part of the \threadgroup so as to ensure that the \threadgroups are static.

\begin{theorem}
    The above implementation is \weak \gl-free if $\{T_{S_1}^i, T_{S_2}^i, \cdots , T_{S_k}^i \}$ implements HarrisListUnsorted  where the groups are $\{\{T_{C_1}, T_{S_1}^1, T_{S_2}^1, \cdots, T_{S_k}^1 \}, \{T_{C_2}, T_{S_1}^2, T_{S_2}^2, \cdots , T_{S_2}^k \}, \cdots \}$.
\end{theorem}

\begin{theorem}
    The above implementation is not \weak \gl-free if $\{T_{S_1}^i, T_{S_2}^i, \cdots , T_{S_k}^i \}$ implements SimpleListUnsorted  where the groups are $\{\{T_{C_1}, T_{S_1}^1, T_{S_2}^1, \cdots , T_{S_k}^1 \}, \{T_{C_2}, T_{S_1}^2, T_{S_2}^2, \cdots T_{S_k}^2 \}, \cdots \}$.

\end{theorem}

\section{Between \Strong and \Weak \gl-freedom/\gw-freedom}
\label{sec:quorum}

While the definitions of \strong and \weak \gl-freedom (respectively \gw-freedom) are identical to the original definitions\cite{on-nature-of-progress}, when \threadgroup consists of 1 thread, they differ significantly in terms of what happens if the \threadgroup contains more than 1 thread. For example, wait-freedom requires that if a thread executes in isolation, it must complete its task in a finite number of steps. This is true for \strong \gw-freedom. A thread is guaranteed to complete its operation even if none of the other threads in its group execute. However, this is not true of \weak \gw-freedom. A thread may not be able to complete its operation in \gw-freedom if even one of the threads in its group does not execute.

The question, therefore, is whether there is a version of \gl-freedom and \gw-freedom that can fall between \strong and \weak \gl-freedom and \gw-freedom. Specifically, is it possible to strengthen \weak \gl-freedom (or \weak \gw-freedom) while avoiding the impossibility result of Theorem \ref{thm:nostrong}. 

The idea of this \intermediate \gl-freedom partitions the \threadgroup $G$ into a set of \quorums, $Q_1, Q_2. \cdots$ such that $G=\bigcup Q_i$. The \threadgroup $G$ is responsible for completing the assigned operation. In \intermediate \gl-freedom (respectively, \gw-freedom), we want to guarantee that as long as one of the \quorums runs infinitely often, the task would be completed. In some sense, this is similar to adding fault-tolerance to initial mutual exclusion algorithms (e.g., \cite{DBLP:journals/cacm/RicartA81}) that failed if even one machine failed, to algorithms \cite{DBLP:journals/tocs/Maekawa85} that ensure correctness as long as at least a quorum of nodes is alive.  

With such \intermediate \gl-freedom, even if some threads in a group do not execute infinitely often, it would be possible that the operation would complete. In other words, the notion of an active group here would only require that one of the \quorums is active rather than requiring that all threads in $G$ are active. 

One instance where \intermediate \gl-freedom can be applied is when there are multiple threads on a single machine. The purpose of the multiple threads is to support expedited performance via concurrent execution. However, if one of the threads fails, other threads could complete the task. 
This technique is often used in concurrent algorithms where threads perform idempotent tasks thereby providing the potential for improved performance without causing failure if a thread blocks. 
%Examples of such instances can be found in \cite{?} where threads perform idempotent tasks to ensure that if one or more threads continue to execute, the operation will be completed. 
In other words, if there are two machines, say $A$ and $B$ with two threads each ($T_{A_1}$, $T_{A_2}$, $T_{B_1}$, $T_{B_2}$) then the relevant \quorums would be $\{ T_{A_1}, T_{B_1} \}, \{ T_{A_1}, T_{B_2} \}, \{ T_{A_2}, T_{B_1} \},$ and $ \{ T_{A_2}, T_{B_2} \}$. As long as one of the quorums of threads is active, the task will be completed. 

With this extension, the operation would be completed as long as one of the quorums in the group is active. Note that the quorum may be machine-based as shown in the example above. It could also be permission based where the quorum has the necessary permissions to complete the task. The notion of quorum used in above discussion is entirely abstract. 

\section{Existing Implementations of \Weak \gl-freedom and \Weak \gw-freedom and Open Problems}
\label{sec:existing}

Our existing implementations of \weak \gl-free data structures were built under extremely weak assumptions: eventual delivery of messages that are sent infinitely often. Under these constraints, busy waiting was the only universally available mechanism for progress, since it required no operating system support and could be executed on any machine. 
%The central objective of that design was to satisfy condition 3(b) of Definition \ref{def:validgroups}, ensuring that progress was preserved even under minimal system assumptions. 

In this section, we show that there already exist several mechanisms that achieve comparable progress guarantees without relying on busy waiting. One such approach  is to combine ZeroMQ \cite{zeromq} with RDMA. We discuss this in Section \ref{sec:glfreewithzeromq}. Then, in Section \ref{sec:aptprimitive}, we discuss the proposed \APT primitive that will simplify the design of future \gl-free and \gw-free algorithms. In Section \ref{sec:limitations}, we discuss some limitations and open problems in the context of \gl-free and \gw-free algorithms.
% \sandeepnote{CheckAtENd}

\subsection{Implementing \gl-free algorithms with ZeroMQ and RDMA}
\label{sec:glfreewithzeromq}
Although ZeroMQ exposes several messaging patterns (pub/sub, req/rep, push/pull, and router/dealer), the push/pull pattern aligns most closely with the examples developed in Sections \ref{sec:singlemachine}, \ref{sec:RPC}, and \ref{sec:dynamicocncurrency}. In the push/pull model, the sender maintains a dedicated outgoing queue for each connected peer and distributes messages in a round robin fashion. Message delivery is retried unless the peer disconnects. This behavior matches precisely the sender side assumptions formalized in Section \ref{sec:RPC}.

By default, ZeroMQ relies on the kernel’s networking stack to transmit messages. When paired with RDMA, however, the data path changes fundamentally: message payloads can be written directly into remote memory without kernel involvement. For the receiver, the RDMA delivered data is placed into the appropriate queue and surfaced as an event to the thread awaiting that message. This mirrors the structure of our own receiver side implementation.

Consequently, the high level structure of this approach can be realized directly using ZeroMQ atop an RDMA transport, yielding an execution path that bypasses kernel scheduling entirely. This ensures that the transition of a thread from \blocked to \executable is not influenced by the executions of unrelated threads in the system.

Note that RDMA either allows an interrupt to be generated at the remote OS or permits polled CQ (Completion Queue). With either of these approaches, it would not be necessary for threads in \blocked state to busy-wait; they can be scheduled only when they need to while still preserving the \gl-free property of algorithms discussed in Sections \ref{sec:singlemachine}, \ref{sec:RPC} and \ref{sec:dynamicocncurrency}.  

\subsection{Mutual Exclusion: CAS primitive :: Network Communication : \APT primitive}
\label{sec:aptprimitive}

Early work in concurrent computing adopted mutual exclusion as the standard mechanism for regulating access to shared objects, ensuring that only one process could enter a critical section at any time. However, a principal disadvantage of this was that once a process acquires the lock, it can impede the progress, intentionally or due to issues such as page faults, of all others.

Primitives such as test‑and‑set, compare‑and‑swap (CAS), and fetch‑and‑add eliminate such delays by reducing it to a minimal hardware‑enforced critical section. For example, the semantics of CAS(a, b, c) assumes the ability to perform the atomic test‑and‑update sequence as though executed with exclusive access. 
Essentially, CAS limits the critical section to an atomic compare‑and‑update and guarantees that no thread can fail or stall within that atomic step.
%\newtext{Essentially, CAS limits critical section to atomic-test-and-update but guarantees that no thread can fail during it. }
%With this modification, a thread may still fail while inside a critical section, but it cannot remain fail inside a CAS operation.

The role of shared‑memory access in concurrent computing is closely analogous to the role of interprocess and interprocessor communication in distributed systems. While hardware primitives such as CAS provide a mechanism that is standardized today, no comparably standardized primitive exists for communication. We propose that an operation of the form \textit{ActivatePartnerThread (\APT) }  (the principle behind Condition 3(b)), which reliably awakens a designated partner thread on the same or a remote machine, can play this role. As shown in Section \ref{sec:glfreewithzeromq}, this functionality can be realized with only user-level code (ZeroMQ) and some support from hardware (RDMA).  Hence, we expect \APT to be one of the primitives that can assist in building \gl-free or \gw-free algorithms. Furthermore, if a protocol is \gl-free, it needs to utilize a primitive that can be built to support \APT. Overall, this implies that \APT is one of the minimal primitives to extend the notion of lock-freedom in the context of distributed systems. Just as it is immaterial how CAS is implemented in the hardware, it is immaterial how \APT is implemented (e.g., the remote thread is woken up or remote thread is created on demand) as long as the partner thread changes to \executable state (if not already in \executable state) in a finite number of steps. 

Since HarrisList is known to outperform SimpleList \cite{Harris2001LinkedList}, it follows directly that any protected, multi‑machine, or distributed variant built atop HarrisList will likewise outperform the corresponding variant built on SimpleList. 

%\sandeepnote{I think  section 10.3 would need to be moved to appendix. DO not see a choice}
\subsection{Limitations in Creating \gl-free and \gw-free Algorithms and Open Problems}
\label{sec:limitations}
We showed that HarrisList can be extended to \gl-free protected, multi-machine or partitioned list. The same idea can be applied to various data structures. However, if the partitioning is dynamic or the algorithms use primitives such as 2-word-CAS, it may introduce challenges in creating \gl-free algorithms. For example, if the list is being dynamically load-balanced then such load-balancing may interfere with the client operation thereby making it not \gl-free. In this context, we note that in \cite{DBLP:journals/corr/abs-2510-06387}, authors have presented a dynamically re-balanced list that is \gl-free even if a part of the list is being moved while the operation is ongoing. However, this approach is limited to the list data structure and may not be generalizable. 
Likewise, if one is using 2-word-CAS \cite{useDCAS}, it can be implemented with
% transformed into a
1-word-CAS using the technique in \cite{MultiCAS} in a single-machine environment. However, this transformation relies on the fact that the words are on the same machine. Whether this problem can be solved in multiple machines is an open problem.

\section{Related Work}
\label{sec:related}
% \spcmnt{I think this section should come before Conclusion \secref{conclusion}}
% \sandeepnote{I put it here so the main part is in 10 pages. I want conclusion to be in 10 pages.}

Wait-freedom and lock-freedom have been examined in two primary contexts: the data structure context, where each thread independently executes its own operation and expects a result, and the common task context, where threads cooperate to accomplish a shared objective, such as computing a spanning tree \cite{BaderCong:MST:JPDC:2006}. Our work builds on the former, extending it to scenarios in which a task must be completed collectively by a group of threads.
While the latter also involves multiple threads, there is a crucial distinction between that setting and ours. In Bader \& Cong MST algorithm \cite{BaderCong:MST:JPDC:2006}, threads are employed merely to accelerate task completion—each thread is individually capable of finishing the task. In contrast, our approach requires genuine group computation, since no single thread possesses the ability to complete the task alone.

Wait-free (respectively, lock-free) computation ensures that every thread (some thread) will make progress even if some other threads are slow/infinitely delayed. This ensures minimal progress for the system. Also, wait-free computation guarantees progress in the worst case. For this reason, various algorithms have been designed to build wait-free/lock-free data structures, to identify universal constructions for wait-free/lock-free computation, and other techniques to improve their performance. 

In \cite{Kogan2012FastPath}, authors present a fast-path-slow-path design pattern to implement wait-free data structures and apply it in the context of queues and linked list. In \cite{Nikolaev2022wCQ}, authors have proposed a wait-free queue that not only guarantees a bound on execution time but bounds memory usage. It presents a circular queue design that achieves similar wait-freedom while maintaining bounded memory. In \cite{ Kogan2018NVMQueue}, the authors propose a lock-free queue for nonvolatile memory and address challenges in crash resilience and concurrent durability. Other lock-free data structures include linked lists, skip lists, hash tables, queues, priority queues, etc. (\cite{ MichaelScott1996Queue, Harris2001LinkedList, SundellTsigas2005SkiplistPQ, ShalevShavit2004HashTable}).

Many wait-free/lock-free algorithms rely on compare-and-swap operation to ensure wait-free design. In \cite{BurdenOfPast2021}, the authors have identified limitations of this compare-and-swap in infinite model, where the space complexity is constrained by the number of active threads. It provides a tunable construction that balances space complexity with progress guarantees. 

Consensus objects \cite{herlihy1991wait} are building blocks for wait-free concurrent objects. This was extended to ($y$,$x$) live objects in \cite{AsymmetricProgress}, where $y$ out of $n$ total processes have wait-free progress and $y-x$ have obstruction-free progress and how such objects can implement a consensus object using arbiters for process groups. The process groups of this work are fundamentally different from group computation model in this paper. It is later shown in \cite{AdversariesIntoFriends} that the liveness in \cite{AsymmetricProgress} can be modeled using adversaries. 

% Since memory reclamation can create a bottleneck in wait-free computation, several studies have proposed hybrid techniques that combine hazard pointers, epoch-based reclamation, and reference counting to improve throughput without compromising wait-freedom (\cite{Nikolaev2022wCQ, Nikolaev2024Crystalline, TurnQueue2021, Nhan2021Thesis}).

Existing algorithms are tailored for single-machine, multi-core environments, where the thread initiating an operation is solely responsible for completing it. In contrast, our work explores the extension of these definitions to distributed, multi-machine settings, where inter-thread coordination across machines is essential and, hence, the current definitions are not applicable. We broaden the definitions of wait-free and lock-free computation to accommodate this distributed context while preserving the semantic guarantees established in single-machine environments. If this happens, we should classify the protocol to be both \gl-free (or \gw-free) and tolerant to crash faults.

\section{Conclusion}
\label{sec:conclusion}

We extend the definitions of wait-freedom and lock-freedom to encompass computations that require collaboration among multiple threads. In these scenarios, a single thread \textit{cannot} complete the task independently due to constraints such as permissions, multi-machine environments, or NUMA architectures.
We presented two versions of this extension: \strong \gl/\gw-freedom and \weak \gl/\gw-freedom. While the former is desirable, it is impossible for most practical scenarios. We showed that \weak \gl/\gw-freedom  coincides with the original formulation in \cite{herlihy1991wait,on-nature-of-progress} when a group consists of a single thread. 
We presented three examples: protected list, multi-machine list and distributed partitioned list. We showed that obtaining them with the use of Harris List \cite{Harris2001LinkedList} results in a \gl-free list while using the SimpleList causes it to be not \gl-free. 

In concurrent computing, both HarrisList and SimpleList are deadlock-free. However, lock-freedom provides a nuance between these two deadlock-free lists whereby HarrisList is lock-free while the SimpleList is not. 
However, when extended to protected environment, multi-machine environment or partitioned distributed list, both HarrisList and SimpleList are only deadlock-free. Neither of them is lock-free. Our extension of lock-freedom to \gl-freedom preserves the nuance of these two lists whereby the extension of HarrisList is \gl-free while the extension of SimpleList is not. 

Our work differs from those where a group of threads are working together to perform a common task (e.g., build a spanning tree) \cite{BaderCong:MST:JPDC:2006}. Here, one thread is capable of completing the entire task if others fail; multi-threaded environment is only for speedup. By contrast, in our setup, threads must collaborate to complete the task, as a thread cannot do it on its own.

This work opens several promising directions for future research. One avenue involves developing new algorithms that uphold wait-freedom or lock-freedom in multi-machine environments by extending designs originally intended for single-machine systems. This is a nontrivial challenge, as single-machine algorithms often rely on specialized instructions (e.g., compare-and-swap) that are not available across distributed systems. Another important direction is benchmarking: while wait-free and lock-free algorithms are known to outperform locking mechanisms under high contention, it is crucial to identify specific scenarios where these advantages are most pronounced. Additionally, our algorithms relied on ActivatePartnerThread as a way to design the algorithms. We showed that this primitive can be implemented with current hardware (RDMA) and user-level code (ZeroMQ). Identifying whether other primitives could be easier to implement in hardware or easier to write \gl-free or \gw-free algorithms is an open problem.

%Additionally, further exploration is needed in operating system design, particularly in ensuring that Condition 3(b) (Definition \ref{def:validgroups}) is satisfied consistently while maintaining optimal average system performance.

\bibliographystyle{plain}
\bibliography{bibliography}

%\section{What is Extending Domain of a Definition}
\newpage

% \iffalse
\appendix

\section{A Note on Domain Extension}

You ask a child in elementary school for the definition of $a^n$ and they will tell you it is $a$ times $a$ times $a$, $n$ times. When the child grows up, they learn that the domain of $n$ can be broadened so that we can define terms such as $a^{0.5}$ or even more complex terms such as $a^i$. Any domain extension in this manner must ensure that they do not violate existing terms (e.g., the value of $a^2$ cannot change with a new definition) and existing properties (e.g., $a^x*a^y=a^{x+y}$) remain true. 

Given the importance of lock-freedom and wait-freedom in defining progress of operations, in this paper, we want to expand the domain to which it can be applied while ensuring that this extension is backward compatible with the current definition. And, this extended definition allows us to distinguish between different executions that would previously have been only deemed deadlock-free. We argue that this lack of nuance distorts algorithm's critical attributes and, therefore, focus on domain extension for these definitions.

\section{Discussion and Relaxation of Assumptions}
\label{sec:discuss}

In this section, we discuss many of the questions raised by the work and relaxation of the assumptions made in the paper.

\noindent \textit{Can we have dynamic \threadgroups?}

Our approach assumes that \threadgroups remain static. In a client-server model, this means that the server maintains a dedicated thread for each client. If dynamic \threadgroups are introduced, where a limited pool of server threads is shared among multiple clients, the execution becomes more complex. In particular, a server thread must be immediately available when a client issues a request. If not, the request will be queued behind others, violating the principles of \gl-free and \gw-free  computation. We can permit dynamic groups if the group is identified at the moment the operation is invoked and this thread can be assigned in such a way that if the thread assignment to one request is delayed it does not prevent the thread assignment to another request. For example, we need to do this without locking the pool of threads but rather by using some sort of lock-free/wait-free set to represent threads. {However, we can overcome this limitation to some extent by defining a superset of the necessary threads as the \threadgroup, where some threads might never need to be activated. Note that the total number of threads would still need to be finite.}

\vspace*{2mm}
\noindent \textit{Permitting a group to complete multiple tasks } In our definitions, we allowed a \threadgroup to execute only one task at a time. If we want a \threadgroup to execute multiple tasks at a time, it creates a technical problem; the competition between these tasks. For example, consider the case where \threadgroup contains two threads $A$ and $B$ and together they are responsible for $T_1$ (started by $A$) and $T_2$ (started by $B$). It is critical that $A$ and $B$ allocate sufficient processing time to execute the two tasks. Specifically, if they prefer $T_1$ over $T_2$ and $A$ continuously creates tasks $T_1', T_1'', \cdots$, it may prevent $T_2$ from being completed even if $A$ and $B$ execute infinitely often. 
This issue can be addressed by requiring that if a \threadgroup is executing multiple actions then it must execute each of them infinitely often. If this is done then definitions of \gl/\gw-freedom would still apply. We argue that if $A$ and $B$ are executing $T_1$ and $T_2$ infinitely often, we should simply treat it as if there are two threads $A_1$ and $A_2$ for $A$ and two threads $B_1$ and $B_2$ for $B$. With this change the resulting operations will be \gl/\gw-free according to the definition in the paper.

%. This may require analysis to be done offline (after the computation is completed) rather than online. 
\vspace*{2mm}
\noindent \textit{Extension to asynchronous RPC (fire and forget).}

It is possible to extend the RPC mechanism described in Section~\ref{sec:RPC} to support asynchronous (fire-and-forget RPCs). This would be similar to Section \ref{sec:dynamicocncurrency} where multiple servers were contacted simultaneously. However, if the number of pending calls is not known upfront, additional shared memory would need to be created between the \userthread and the NIC thread. The creation of this memory would have been before the operation is invoked, as the creation of shared memory cannot be done without OS intervention.

\vspace*{2mm}
% \sandeepnote{Removing this as we introduce deadlock-freedom in paper}
\iffalse
\noindent \textit{Other properties such as deadlock-freedom and starvation-freedom.}

We did not discuss them because existing approaches already allow us to characterize group computations to be deadlock-free. Those proofs remain valid. What we are able to do is that by studying the interdependency between threads, we can characterize some of them to be \gl-free and \gw-free by observing that the dependencies are between threads that are collaborating to complete the task. 
\fi 

\vspace*{2mm}
\noindent \textit{Asynchronous operations.}

Definition \ref{def:validgroups} requires that when the operation completes, all threads in the \threadgroup must be in \blocked state. This could be changed so that some work can be completed asynchronously after the operation completes. However, if such an approach is desired, there are two concerns that have to be dealt with: First, it is necessary to add a condition that the work done asynchronously must be finite. Second, such an approach may cause unexpected delays to the second operation in the same group because some thread is still executing the asynchronous work from the previous operation. 

\vspace*{2mm}
\noindent \textit{Handling failure of machines.}

Our extension of \intermediate \gl/\gw-freedom permits the possibility of machine failure. Specifically, it only requires a quorum of threads within a group to be active. Therefore, it can handle the situation when a quorum of nodes is active even when some threads fail. A typical example of this would be replicated data where the operation will complete as long as threads of at least one replica execute.

\vspace*{2mm}
\noindent \textit{Failure of threads.}

We note that failure of a thread may prevent the corresponding \threadgroup from completing the task. However, in \gw-free implementation, any active \threadgroup will be able to complete its operation. And, in \gl-free implementation, some active \threadgroup will be able to complete the task. This is discussed in Section \ref{sec:quorum}.

\vspace*{2mm}

\section{Other Alternative Extensions}
\label{sec:otheralternative}

When extending an existing definition, a natural concern is whether the chosen formulation is preferable to other plausible alternatives. Since the space of all potential extensions of lock freedom and wait freedom is not itself a formal mathematical domain, it is impossible to \textit{prove} that our definition is objectively the best. However, we examine several alternatives that a reader might reasonably consider and argue that our formulation provides a stronger and more appropriate extension.

\vspace*{2mm} 
\noindent \textit{
In a client-server model, before the client invokes the operation, it is being executed on the client. Subsequently, it is executed on the server. Hence, one way to define \gl-freedom and \gw-freedom is to utilize the notion of thread migration to extend the definition of lock-freedom and wait-freedom. Conceptually, the entire system can be viewed as a single large machine containing all participating nodes, so a client’s invocation of a server operation corresponds to the client’s thread migrating to the server to continue execution there.}

\vspace*{1mm}
We note that while this approach suffices for the approach in Section \ref{sec:RPC}, it fails to capture the notion of dynamic concurrency in Section \ref{sec:dynamicocncurrency}. Specifically, in the area of thread migration, it is expected that only one thread is executing at a time. By contrast, in \gl-free and \gw-free definition considered in this paper, it is possible that multiple threads can be active in the middle of an operation. This is a desirable option. 

\vspace*{2mm} 
\noindent \textit{
The definition of wait-freedom can be viewed as, `If a thread executes a certain number of steps, then it is guaranteed to complete its operation’. One way to extend it is to say `If a \threadgroup executes a certain number of steps, then it is guaranteed to complete its operation’.}

\vspace*{1mm}

We argue that this is not a desirable definition. As shown in Section \ref{sec:backward}, to utilize such a definition, it is critical that only threads that can make progress are in \executable state. Otherwise, the adversary will assign steps to threads in \blocked state thereby completing the budget that the \threadgroup has. In other words, this prevents the possibility of multiple threads in a \threadgroup utilizing some type of busy-waiting to ensure that they are able to communicate without assistance from the operating system(s). 
By contrast, our definition is more abstract. It permits threads in \threadgroup to communicate via their own mechanism or a mechanism supported by the OS provided that this communication guarantees that thread $i$ can activate thread $j$ in a finite number of steps. 

\vspace*{2mm} 
\noindent \textit{  Yet another option is to model the networking threads explicitly so that we can handle the communication among threads.}

\vspace*{1mm}
This option is reasonable.  {Our approach from abstraction used in requirement 3(b) also permits this.
Specifically, as shown in Section \ref{sec:RPC}, our definition permits RPC communication performed by networking threads. However, networking threads do not allow us to model interprocess communication (e.g., as done in Section \ref{sec:singlemachine}). Additionally, our approach is not tied to a specific implementation. For example, threads in a \threadgroup could communicate via shared file system, distributed shared memory, etc.  
Our approach on the other hand is more generic and extensible to any interrupt handlers of the operating system.

\section{Examples of \gl-free/gw-free Algorithms}
\label{sec:examplesgl}

In this section, we provide a few examples of \gl-free/gw-free algorithms. First, in Section \ref{sec:trivial}, we identify \textit{rudimentary} method to transform a lock-free/wait-free algorithm into corresponding \gl-free/gw-free algorithms. Section \ref{sec:glfreescheduling} discusses a \gl-free solution for the meeting problem in Section \ref{sec:intro}. And, Section \ref{sec:blockchainEG} provides an outline for a \gl-free example in the context of blockchains. 

\subsection{Rudimentary \gl-free/\gw-free Algorithms}
\label{sec:trivial}

Consider a lock-free algorithm for a data structure, e.g., a list \cite{Harris2001LinkedList}. Now, consider the case where this data structure is put on a server and clients are expected to contact the server to access that data structure. In this case, the algorithm is deemed not lock-free since the client thread has to wait for the server thread to complete the task; it will only be considered deadlock-free. However, it will be \gl-free if the server has a separate thread for each client and their communication utilizes mechanism discussed in Section \ref{sec:RPC}. 

Note that if the list operations are performed by locking the list then the resulting list will not be \gl-free. The domain extension in this paper provides this nuance rather than simply calling both implementations to be deadlock-free. 

Note that the discussion in the above paragraph is valid for any lock-free/wait-free data structure. Specifically, if we had started with a lock-free (respectively, wait-free) structure then it will result in a \gl-free (respectively, \gw-free) structure. 

\subsection{\gl-free Implementation for Scheduling Meeting Problem from Section \ref{sec:intro}}
\label{sec:glfreescheduling}

In this section, we outline a \gl-free approach to the meeting scheduling problem introduced in Section \ref{sec:intro}, adopting several simplifying assumptions to streamline the solution. The goal of this section is \textit{only to show the feasibility} of implementing meetings in \gl-free manner.

We assume that the number of principals is finite and fixed at the start of the system. Subsequently, we illustrate the algorithm with Principals A, B, C and D that control their own calendars. However, the algorithm can be generalized to any number of principals. 

We assume that there are $n$ \threadgroups, i.e., $n$ simultaneous meetings can be scheduled at a time. Each \threadgroup contains one thread per principal. It is anticipated that if a meeting involves Principals B and C then only threads involved on those principals will do the work, i.e. threads on A and D will not have to do any work. (We note that some notion of \textit{helping} discussed later may require threads on A and D to do some work.)
We assume that the slots are $s_1, s_2, \cdots$ which are mutually exclusive and exhaustive. The number of slots can be finite or infinite. Each principal maintains some status of that principal for the given slot. The status is either \free (available for booking), \tentativek (tentatively booked by \threadgroup $k$) or \blockedk (booked by \threadgroup $k$). 

When \threadgroup $i$ wants to book a meeting slot, the following rules apply. First, it attempts to book $s_1$. If it concludes that $s_1$ is unavailable, it goes to $s_2$ and so on. To attempt the booking of a slot $s_l$, \threadgroup $i$ will always try to book a slot on A’s calendar (if required) before B’s calendar and so on. So, if the request for booking for \threadgroup $i$ is for principals B and C, then first this \threadgroup will attempt to book a slot on B’s calendar and then C’s calendar. 

When \threadgroup $i$ finds that the calendar status for slot $s_l$ is \free, it marks it as \tentativei (using compare and swap instruction) and asks the thread in the next principal to block slot $s_l$. If it finds that the calendar status is \blockedj where $i \neq j$ then it asks its other threads in its \threadgroup to release any tentative slots they may have booked, as slot $s_l$ is not a viable option. 
Then, it moves onto the next slot, i.e., $s_{l+1}$. If it finds that the status is \tentativei or \blockedi then it asks the thread at the next principal in order to book slot $s_l$. (We will discuss why this could happen later.) If \threadgroup $i$ marks all its desired slots as \tentativei then it asks the members of its \threadgroup to mark it as \blockedi and return the slot to the principal that requested to schedule the meeting. 

Note that the above discussion is only missing the case where \threadgroup $i$ finds that a status is \tentativej where $i\neq j$ for some slot $s_l$. If this happens, \threadgroup $i$ would have to wait to determine if \threadgroup $j$ has completed the entire booking and, hence, slot $s_l$ is no longer available. Or, \threadgroup $j$ will fail to find the required slots for the remaining principals and, hence, will release slot $s_l$ from all calendars. Such waiting is not permitted in \gl-free implementation. Hence, \threadgroup $i$ \textit{helps} \threadgroup $j$. To do so, it identifies the request that \threadgroup $j$ is processing. (We will discuss how to do this in the next paragraph). Then, it will complete the task of booking slot $s_l$ on behalf of \threadgroup $j$. While doing this, if it encounters the status \tentativek where $i\neq k \wedge j \neq k$ then it will try to complete the work of \threadgroup $k$ and so on. Since the principals are ordered, there cannot be cyclic dependencies for any slot. {The key invariant we aim for is to align all \threadgroups competing for a slot $s_l$ to help the \threadgroup that is closest to completing its booking of $s_l$. A \threadgroup $i$ is considered to be closer to booking a slot $s_l$ than another \threadgroup $j$, if it wins the race in setting the tentative value for $s_l$ on a common calendar. A \threadgroup $k$ is said to have begun finalizing a slot $s_l$, if it set the tentative value of $s_l$ in the calendar of the highest key in its request. The remaining \threadgroups then help block the calendars in the request of $k$ and then seek a different slot to book.}

For \threadgroup $i$ to find out the request being processed by \threadgroup $j$, instead of saving \tentativej in the status field, we should store $\langle tentative, j, counter_j, requestset_j \rangle$, where $counter_j$ is a counter that is incremented every time \threadgroup $j$ completes a booking successfully, and $requestset_j$  denotes the set of principals that are involved in the current request. We anticipate that for most systems, this could be stored in a single integer. If not, we can use a descriptor pointer\footnote{{Descriptors typically describe the task to be done and its current status. Here, the set of calendars becomes the task and since we book in the increasing order of calendar, the last calendar with the tentative descriptor of this task describes its current status.} } (much like the ones used in multi-word compare-and-swap \cite{MultiCAS} and lock-free locks \cite{lockFreeLocksRevisited}), whose location contains the tuple that we would like to store. Now irrespective of whether \threadgroup $j$ getting CPU cycles to execute its remaining steps of booking $s_l$, any competing thread that would help see this through - either complete booking $s_l$ for thread j, or notice conflict and remove thread j from the \tentativej state.  After this help, \threadgroup $i$ proceeds to book $s_{l+1}$ if the help led to a successful booking for \threadgroup $j$, proceeds to book $s_{l}$ otherwise.

%\raaghavnote{\textbf{Bold: Please adjust the invariant argument here, which I feel you know best:}} 
When there is no contention (no conflicting slots being booked), naturally no \threadgroups wait for each other. When there is contention, note that the booking task of at least one \threadgroup (the \threadgroup that was closest to finishing its booking) completes either by the assigned \threadgroup of the booking task or by one of the helpers. This is because, the \threadgroup $i$ that booked the slot tentatively in the highest calendar key $C^i_{highest}$  so far, will only see free state of the slot in the remaining calendars that it will need to book. The remaining \threadgroups competing for this slot, will eventually notice this tentative slot of the highest calendar $C^i_{highest}$ and help complete the operation by helping. Thus, the implementation is \gl-free. However, this implementation is not lock-free in the traditional sense, as each principal has a ``current calendar'' that can only be accessed by a certain thread (such as thread A1 and not B1 for Calendar A) ,i.e., if only threads for Calendar B are given computational steps, it is possible that no operation gets completed.
%We can achieve better performance \raaghavnote{by using lock-free locks\cite{lockFreeLocksRevisited} on each slot that is to be booked tentatively for this purpose,} but we are not focusing on that since the goal of this section is to only illustrate feasibility \gl-free implementation rather than the most efficient implementation.  
We note that our approach is similar to using lock-free locks \cite{lockFreeLocksRevisited} and the methodology in \cite{lockFreeLocksRevisited} will {both simplify the presentation and may even offer better performance.} However, we are describing the algorithm from scratch to keep the paper self-contained, as our goal is only to illustrate the feasibility  of \gl-free implementation.

\subsection{Additional Example for Inability of a Single Thread to Complete the Operation}
\label{sec:blockchainEG}

So far, we have assumed that all systems are asynchronous, i.e., there was no notion of time. However, the same framework can also be applied to settings that include synchronous systems where we have a notion of a common time and, hence, threads can utilize timeout mechanisms, e.g., blockchains. 

Consider, for example, the abstraction model of blockchains introduced by Maurice Herlihy in the context of atomic cross-chain swaps, where individual blockchain platforms are treated as independent systems~\cite{10.1145/3212734.3212736}. The solution proposed in \cite{10.1145/3212734.3212736} considers possible scenarios where the participants are dishonest, i.e., they fail to complete their own part of the task only to find that they lose money when they do. For the sake of simplicity, we only focus on the part where each participant completes its task for performing the actual swap.

An atomic cross-chain swap is a canonical example of a task that cannot be safely executed by a single system in isolation and therefore requires coordination between multiple independent systems \cite{10.1145/3212734.3212736}. In other words, in these systems, no single thread can complete the operation on its own. Cooperation across systems (and, therefore, among several threads) is a must for completing the operation. A blockchain is a distributed ledger maintained by a network of nodes; different blockchains, such as Bitcoin and Ethereum, operate as completely separate systems with no inherent ability to observe or verify each other’s state. 

Now consider a transaction that spans both systems. Suppose Alice wants to sell Ether (Ethereum’s currency) in exchange for Bitcoin from Bob. Conceptually, this is a single deal, but technically it requires two separate transactions:
\begin{enumerate}
    \item Alice sends Ether on the Ethereum blockchain.
    \item Bob sends Bitcoin on the Bitcoin blockchain.
\end{enumerate}
Either both transactions must occur, or neither should. If only one transaction succeeds, one party could lose money. The Bitcoin network cannot see or verify what is happening on the Ethereum network, and vice versa. Each network alone cannot safely complete the cross-chain exchange.

A range of solutions currently support cross-chain swaps across multiple blockchains. However, the literature does not clearly distinguish between interoperability mechanisms that allow a single wallet (e.g., Alice’s wallet) to execute parallel cross-chain and on-chain transactions and those that require the wallet to be locked for the duration of a cross-chain operation. 

For example, consider a system where Alice wants to perform a swap with Bob and simultaneously pay Carol for a cup of coffee. An implementation where Alice's wallet is locked for currency swap with Bob, she would not be able to perform the transaction with Carol. By contrast, if only a small part of the currency involved in the swap is locked then transaction with Carol can succeed. 

Here, we can assume that there are two \threadgroups $A1$ and $B1$ that perform the currency swap between Alice and Bob and another \threadgroup $A2$ and $C1$ that perform payment from Alice to Carol. 
In the implementation where the entire wallet is locked, $A2$ has to wait for $A1$ (or vice versa). Hence, this implementation cannot be \gl-free or \gw-free. By contrast, if only a limited funds are blocked, $A1$ does not need to wait for $A2$ and vice versa. This allows the solution to be \gw-free (assuming locks are not used in other parts of the transfer). 

We note that in either case, these operations are not lock-free or wait-free as $A1$ must wait for $B1$ to complete its task and $A2$ must wait for $C1$ to complete its task. 
%In other words, it would be possible to show the solution that locks the wallet to be not \gw-free. By contrast, the solution that locks limited currency \textit{could be} \gw-free. 

We note that while currency swap and payment could be handled in \gl-free or \gw-free manner, there \textit{may be} limitations on what can be achieved with \gl-free or \gw-free guarantees. For example, if the system supported currency swap and percentage of asset transfer (where Alice wants to transfer 10\% of her assets to Carol), it is unclear if these two operations could be performed in a \gl-free or \gw-free manner. This issue is possibly because of limitations of achieving {non-blocking} serializability with transactions \cite{linearizability}. This example also suggests some limitations under which \gl-free or \gw-free algorithms can be designed. 

\begin{figure}[htbp]
\centering
\resizebox{0.95\linewidth}{!}{%
\begin{tikzpicture}[
    system/.style={
        draw, rounded corners,
        minimum width=5.4cm, minimum height=2.6cm,
        align=center, inner sep=8pt, ultra thick, font=\Large
    },
    tx/.style={
        draw, rounded corners,
        minimum width=4.2cm, minimum height=1.0cm,
        align=center, inner sep=5pt,
        fill=red!10, ultra thick, font=\Large
    },
    arrow/.style={->, ultra thick},
    dblarrow/.style={<->, ultra thick},
    note/.style={font=\Large, align=center}
]

\node[system, fill=blue!20] (eth) at (-5,0)
{\textbf{Ethereum Network}\\(Independent System)};

\node[system, fill=green!20] (btc) at (5,0)
{\textbf{Bitcoin Network}\\(Independent System)};

\node[tx] (ethTx) at (-5,-0.5)
{\textbf{Transaction 1:}\\Alice sends Ether};

\node[tx] (btcTx) at (5,-0.5)
{\textbf{Transaction 2:}\\Bob sends Bitcoin};

\node[note] at (-5,1.9)
{Cannot observe or verify\\Bitcoin transaction};

\node[note] at (5,1.9)
{Cannot observe or verify\\Ethereum transaction};

\draw[dblarrow] (-1.9,0.4) -- (1.9,0.4);
\node[note] at (0,0.8) {Atomic cross-chain swap};

\node at (-5,-2.9) {\tiny
\begin{tikzpicture}[scale=0.5]
\draw (0,0) circle (0.4);
\draw (0,-0.4) -- (0,-1.2);
\draw (-0.6,-0.8) -- (0.6,-0.8);
\draw (-0.4,-1.6) -- (0,-1.2);
\draw (0.4,-1.6) -- (0,-1.2);
\end{tikzpicture}
};
\node[below=0.9mm] at (-6,-2.6) {ALICE};

\node at (5,-2.9) {\tiny
\begin{tikzpicture}[scale=0.5]
\draw (0,0) circle (0.4);
\draw (0,-0.4) -- (0,-1.2);
\draw (-0.6,-0.8) -- (0.6,-0.8);
\draw (-0.4,-1.6) -- (0,-1.2);
\draw (0.4,-1.6) -- (0,-1.2);
\end{tikzpicture}
};
\node[below=1mm] at (5.9,-2.6) {BOB};

\draw[arrow] (-5,-2.3) -- (ethTx.south);
\draw[arrow] (5,-2.3) -- (btcTx.south);

\node[note] at (0,-3.6)
{Either both transactions occur, or neither occurs\\
(Otherwise, one party may lose funds)};

\end{tikzpicture}
}
\caption{Atomic cross-chain swap as a real-world example where no single system suffices.}
\end{figure}
% \fi
%\iffalse 
%\fi
\end{document}

%% file: macros.tex
\newcommand{\cmnt}[1]{}
\newcommand{\ignore}[1]{}
\newcommand{\remove}[1]{}

\newcommand{\tentativei}{\ensuremath{tentative_i}\xspace}
\newcommand{\tentativej}{\ensuremath{tentative_j}\xspace}
\newcommand{\tentativek}{\ensuremath{tentative_k}\xspace}
\newcommand{\blockedi}{\ensuremath{blocked_i}\xspace}
\newcommand{\blockedj}{\ensuremath{blocked_j}\xspace}
\newcommand{\blockedk}{\ensuremath{blocked_k}\xspace}
\newcommand{\free}{\textit{free}\space}

\newtheorem{definition}{Definition}
\newtheorem{observation}{Observation}

\definecolor{raaghavColor}{rgb}{0.3, 0, 0.6}
\definecolor{sandeepColor}{rgb}{0.6, 0, 0.3}
\definecolor{spColor}{rgb}{0.3, 0, 0.3}

\newcommand{\raaghavnote}[1]{{\color{raaghavColor}#1}}
\newcommand{\sandeepnote}[1]{{\color{sandeepColor}#1}}

\newcommand{\executable}{\ensuremath{EX}\xspace}

\newcommand{\blocked}{\ensuremath{BL}\xspace}

\newcommand{\userthread}{application thread\xspace}
\newcommand{\userthreads}{application threads\xspace}

\newcommand{\Userthreads}{Application threads\xspace}
\newcommand{\systemthread}{background thread\xspace}
\newcommand{\systemthreads}{background threads\xspace}

\newcommand{\Systemthreads}{Background threads\xspace}
\newcommand{\threadgroup}{\textsf{thread-group}\xspace}
\newcommand{\threadgroups}{\textsf{thread-groups}\xspace}

\newcommand{\ThreadGroup}{\textsf{Thread-Group}\xspace}
\newcommand{\ThreadGroups}{\textsf{Thread-Groups}\xspace}

\newcommand{\gl}{gl\xspace}
\newcommand{\gw}{gw\xspace}

\newcommand{\Weak}{Weak\xspace}
\newcommand{\Strong}{Strong\xspace}

\newcommand{\weak}{weak\xspace}
\newcommand{\strong}{strong\xspace}
\newcommand{\weakactive}{\ensuremath{\forall}-active\xspace}
\newcommand{\strongactive}{\ensuremath{\exists}-active\xspace}

\newcommand{\WeakActive}{\ensuremath{\forall}-Active\xspace}
\newcommand{\StrongActive}{\ensuremath{\exists}-Active\xspace}
\newcommand{\APT}{APT\xspace}

\newcommand{\intermediate}{intermediate\xspace}

\newcommand{\quorums}{quorums\xspace}

%% file: committeexample.tex
\begin{figure}[htbp]
\centering
\resizebox{\linewidth}{!}{%
\begin{tikzpicture}[
    proc/.style={draw, ultra thick, minimum width=1.3cm, minimum height=5.6cm},
    band/.style={draw, ultra thick, minimum height=1.2cm},
    label/.style={font=\Huge},
    note/.style={font=\Huge, align=left, text width=6.0cm},
    arrow/.style={->, ultra thick}
]
% X positions
\def\xB{2.6}
\def\xA{0}
\def\xC{5.2}
\def\xD{7.8}

% Process columns
\node[proc] at (\xB,0) {};
\node[proc] at (\xA,0) {};
\node[proc] at (\xC,0) {};
\node[proc] at (\xD,0) {};

% Process labels
\node[label] at (\xA,3.3) {System A};
\node[label] at (\xB,3.3) {System B};
\node[label] at (\xC,3.3) {System C};
\node[label] at (\xD,3.3) {System D};

% Shared bands
\node[band, minimum width=12cm] (topband) at (2.6,1.2) {};
\node[band, minimum width=12cm] (botband) at (5.2,-1.2) {};

% Operation labels
\node[label] at (\xA,1.2) {$A_1$};
\node[label] at (\xB,1.2) {$B_1$};
\node[label] at (\xC,1.2) {$C_1$};
\node[label] at (\xD,1.2) {($D_1$)};

\node[label] at (\xA,-1.2) {$A_2$};
\node[label] at (\xB,-1.2) {($B_2$)};
\node[label] at (\xC,-1.2) {$C_2$};
\node[label] at (\xD,-1.2) {$D_2$};

% Side explanations
\node[note, left=1.0cm of topband] (topnote)
{Threads $A_1$, $B_1$, and $C_1$ collaborate to find a common timeslot and schedule a meeting between principals $A$, $B$, and $C$.};

\node[note, right=1.0cm of botband] (botnote)
{Threads $A_2$, $C_2$, and $D_1$ collaborate to find a common timeslot and schedule a meeting between principals $A$, $C$, and $D$.};

% Arrows
\draw[arrow] (topnote.east) -- (topband.west);
\draw[arrow] (botnote.west) -- (botband.east);

\end{tikzpicture}
}
\caption{\ThreadGroup coordination for meeting scheduling. 
\ThreadGroup $\{A_1,B_1,C_1\}$ and $\{A_2,C_2,D_2\}$ collaborate to schedule meetings.}
% Appendix \ref{sec:glfreescheduling} adds $D_1$ and $B_2$ and identifies their role.}
\label{fig:thread-group-coordination}
\vspace*{-5mm}
\end{figure}